\documentclass[twocolumn]{aastex63}

\received{ --- }
\revised{ --- }
\accepted{ --- }
\submitjournal{ApJ}

\shorttitle{Effects of Magnetic Field Loops}
\shortauthors{Garain et al.}

\begin{document}

\title{Effects of Magnetic Field Loops on the Dynamics of Advective Accretion Flows and Jets around a Schwarzschild Blackhole}

\correspondingauthor{Sudip K. Garain}
\email{sgarain@nd.edu, sgarain@kasi.re.kr}

\author{Sudip K. Garain}
\affiliation{Department of Physics,
University of Notre Dame,
Notre Dame, IN 46556, USA}
\affiliation{Korea Astronomy \& Space Science Institute,
776 Daedeokdae-ro, Yuseong-gu, Daejeon 34055, Korea}
%
\author{Dinshaw S. Balsara}
\affiliation{Department of Physics,
University of Notre Dame, 
Notre Dame, IN 46556, USA}
\author{Sandip K. Chakrabarti}
\affiliation{Indian Center for Space Physics,
43 Chalantika, Garia St. Rd.,
Kolkata, 700084, India}
\author{Jinho Kim}
\affiliation{Department of Physics,
University of Notre Dame,
Notre Dame, IN 46556, USA}
\affiliation{Korea Astronomy \& Space Science Institute, 
776 Daedeokdae-ro, Yuseong-gu, Daejeon 34055, Korea}
%





\begin{abstract}

Magnetic fields advected along with low angular momentum accretion flows predominantly
become toroidal due to the strong azimuthal velocity close to a black hole.
We study self-consistently the movements of these flux tubes inside an advective disc
and how they dynamically influence the flow. We find that the centrifugal barrier
slows down the radial motion of the flux tubes. In this case, the large magnetic flux tubes with a significant drag
force escape along the vertical axis due to buoyancy. Magnetic pressure rises close to the black hole and
together with the centrifugal force, it combats gravity. The tug-of-war
among these forces causes the centrifugal pressure supported shock to oscillate radially.
We study the effects of successive injection of flux tubes and find how the flux tube could be
trapped inside the disc in regions of highest entropy. Most interestingly, the shock wave
remains at its average location and is not destroyed. We show that the toroidal field loops
contribute significantly to collimate and accelerate the outflows from the centrifugal barrier
and suggest this mechanism to be a way to collimate and accelerate jets.

\end{abstract}

\keywords{accretion, accretion disks --- methods: numerical --- black hole physics ---
magnetohydrodynamics (MHD) --- plasmas --- shock waves}


\section{Introduction}

Magnetic field is ubiquitous in nature. In the context of accretion flows in a binary system containing a black hole, 
it is likely that the companion star may have winds with entangled magnetic fields which are 
accreted by the black hole. As the accretion flow approaches the black hole, the azimuthal velocity of the 
flow increases and in an ideal MHD limit, the spiraling flow stretches the field line and makes 
predominantly toroidal loops. As these loops come very close to the black hole, 
the radial velocity rises and becomes supersonic, stretching the field lines in radial direction. Due to drag
effect, the radial motion of larger flux tubes is slowed down and due to buoyancy effect they may leave the disc 
altogether (\citealt{cd1994} hereafter CD94; \citealt{dsilva1994} hereafter DC94).

In the present paper, our goal is to carry out a self-consistent study of the dynamic behavior of these 
magnetic flux tubes inside an advective, low-angular momentum, flow around a Schwarzschild black hole. We also  
study the effects of these flux tubes on the flow. In the literature, qualitative studies of the  
role of the flux tubes in transporting angular momentum and creating possible corona in an accretion flow 
have been discussed in \citet{el1975, galeev1979, coro1981, shibata1990}.
Dynamics of flux tubes in a steady disc has been studied in detail in CD94 and DC94 where the flux tubes 
have been assumed not to influence the dynamics of the flow. The forces such as the magnetic buoyancy, tension, 
are included in those calculations.
They demonstrated that if the flux tubes are very thin, then 
they can reach very close to the black hole, otherwise they would 
leave the low angular momentum thick disc along the funnel wall.  

It is long speculated that the funnels in thick accretion discs \citep{lynden1978,pw1980}
may be a site to collimate the outflows and jets observed 
in Active galactic nuclei and micro-quasars. \citet{eck1985} discussed 
radiative acceleration of jets from the funnel and found a maximum velocity of about 
$0.3c$, where $c$ is the velocity of light. \citet{fukue1982,chakraba1986} discussed 
acceleration of the jets using thermal and hydrodynamical processes. In all these 
cases the matter is supplied by the thick accretion disc, while collimation is done 
by the centrifugally driven vortices or funnel walls along the axis. 
\citet{lovelace1976,bp1982,chakraba1992,camen1989, heyv1989, koide1999}
and others show that magnetic fields may also contribute 
to the collimation of the outflows. \citet{konigl1989} showed that in some region of 
the parameter space it is possible to obtain self-similar \citet{bp1982}
type jets which achieve super-Alfvenic velocity soon after the matter leaves the disc. 
In the recent simulations by general relativistic (GR) magnetohydrodynamic (MHD) codes, 
matter is usually launched from the discs threading poloidal magnetic fields and 
jets are produced \citep{nishi2005, mckinney2006,shafee2008,tchekhov2011}.
The nature of the source of such aligned large 
scale magnetic field has not been discussed in these works and results are found to be sensitive to 
the initial field configuration.

Since it is difficult to imagine, how, in the absence of poloidal fields of central 
black holes, a disc can have large scale unidirectional poloidal fields, we are 
motivated to take up a realistic problem where such large scale fields are not 
required in order to produce well collimated and accelerated jets, provided 
matter is injected from an advective disc. A low angular momentum advective flow moves 
rapidly and thus feels the centrifugal barrier very close to the black hole. The outer 
boundary of matter piled up at this barrier is the shock-front and post-shock region
downstream behaves as a thick disc as shown more than 
two decades ago \citep{mlc1994} using a hydrodynamic simulation code. 
This thick disk is believed to produce the observed hard radiation from the accretion disk 
(\citealt{chakraba1997} and references therein) 
and the oscillation of the centrifugal pressure supported 
shock boundary may cause the quasi-periodic oscillations \citep{msc1996,cam2004,ggc2014} 
observed in light curves of several black hole binaries as revealed by the power-density spectra.
Thus, efforts are on to include magnetic field to this advective flow which naturally 
produces thick discs as used in earlier days. After the work of CD94 and DC94 where the 
disk variables were pre-determined, a recent work by \citet{deb2017}
studied the motion of flux tubes in a {\it hydrodynamic} time dependent flow showing 
that the injection of a single flux tube increased the collimation of the jet as long as it 
did not escape.
However, there was no attempt to study MHD flows or the effects of the fields on 
the dynamics of the accretion flow. In the present paper, we remove these deficiencies 
and answer the followings questions: (a) Can the toroidal field lines escape due to 
buoyancy? (b) Are they capable of collimating and accelerating jets? (b) Are the 
standing shocks in an advective flow stable under these axisymmetric flux tubes? 
(c) Can there be a steady corona in an advective disc which may inverse Comptonize the flow? 
(d) Is there a way to anchor fields inside an advective disc? (e) Are there both steady 
jets and episodically ejected blobby jets in a magnetized flow? Of course, every 
numerical simulation also produces new results which are not anticipated before and 
the present one no exception. We will discuss them and their implications in the final Section.

In the next Section, we shall present numerical method used in our simulations. 
In Section 3, we present the results and finally in Section 4, we present the concluding remarks.

In this paper, we choose $r_g=2GM_{\rm{bh}}/c^2$ as the unit of distance,
$r_g c$ as unit of angular momentum, and $r_g/c$ as unit of time.
Here, $G$ is the gravitational constant and $M_{\rm{bh}}$ is the mass of
the black hole. In addition to these, we choose the geometric units $2G=M_{\rm{bh}}=c=1$.
Thus $r_g=1$, and angular momentum and time are measured in dimensionless units.

\section{Numerical Methods}

\label{sec:sec1}

In order to study the magnetized sub-Keplerian advective disc, we solve the 
non-relativistic ideal MHD equations in cylindrical coordinates. A realistic 
disc is three dimensional. However, assuming axisymmetry, we simplify the problem 
and solve these equations numerically in the R-Z plane. The full set of equations can be
written in the conservation form as follows \citep{ryu1995jkas,bal2004}
\begin{equation}
\frac{\partial \rho}{\partial t} +
\nabla \cdot \left( \rho {\mathbf v}\right) = 0
\end{equation}

\begin{equation}\label{momeqn}
\frac{\partial \left( \rho {\mathbf v}\right)}{\partial t} +
\nabla \cdot \left(\rho \mathbf {v\otimes v} + \left( P + \frac{{\mathbf B}^2}{8\pi}\right) {\mathbf I} 
  - \frac{\mathbf {B\otimes B}}{4\pi} \right) = -\rho {\mathbf g}
\end{equation}

\begin{equation}
\frac{\partial E}{\partial t} +
\nabla \cdot \left( \left( E + P + \frac{{\mathbf B}^2}{8\pi} \right){\mathbf v} - \frac{\mathbf {(B\cdot v)B}}{4\pi}\right) = -\rho {\mathbf v} \cdot {\mathbf g}
\end{equation}

\begin{equation}
\frac{\partial {\mathbf B}}{\partial t} - \nabla \times \left( \mathbf{v \times B} \right)=0
\end{equation}

Here, $\rho$ is density, $P$ is the thermal pressure, ${\mathbf v}$ is the velocity,
${\mathbf B}$ is the magnetic field, ${\mathbf I}$ is the identity tensor, 
${\mathbf g}$ is the gravitational acceleration and
$$E = \frac{1}{2}\rho v^2 + \frac{P}{\gamma -1} + \frac{B^2}{8\pi}$$ is the energy.

We use the second order accurate RIEMANN code 
\citep{bal1998a,bal1998b,bal2004,bal2009,bal_spi1999a,bal_spi1999b,brdm2009,bal2013}
to solve the ideal MHD equations. Spatial reconstruction has been carried out using MC limiter
and HLL Riemann solver has been used for flux calculations. In addition, multi-dimensional Riemann
solver has been used for edge-averaged electric field evaluations. The matching time accuracy
has been achieved following predictor-corrector step. To model the initial injection, we assume an 
accretion flow of a gas around a non-rotating black hole located at the 
center of the cylindrical coordinate system. The gravitational potential of the 
central black hole is modeled using the pseudo-Newtonian potential given by 
\citet{pw1980},
$$
\psi(r)=-\frac{1}{2\left(r-1\right)},
$$
where, $r=\sqrt{R^2+Z^2}$. We also assume a polytropic equation of state 
for the accreting gas, $P=K\rho^\gamma$, where, $\gamma = 4/3$ is the adiabatic constant.
$K$ is the measure of entropy and is allowed to change inside the disc.

Since we are interested in the region very close to the central black hole, 
the simulations are performed on a $\left[2:100\right]{\rm r_g}\times\left[-50:50\right]{\rm r_g}$
computational domain in the R-Z plane using a uniformly spaced $1024\times 1024$
zone mesh. Incoming matter enters the computational domain through the right radial 
boundary at ${\rm R_{out}=100r_g}$. The incoming matter injected at the right boundary
points towards the central black hole. 
The radial speed $v_{\rm in}=\sqrt{v_R^2+v_Z^2}$ of the incoming matter is the same at all heights. 
We also use the same sound speed $a_s$ at all heights \citep{mrc1996}. 
The values of $v_{\rm in}$ and $a_s$ are calculated following \citet{chakraba1989,chakraba1990}
assuming the vertical equilibrium model as they depend on the specific energy $\epsilon$
and specific angular momentum $\lambda$ of the incoming matter. The interior of the 
computational domain is initialized with a low density given by $\rho_{\rm floor}=10^{-6}$.
The initial low pressure $P_{\rm floor}=\frac{a_s^2 \rho_{\rm floor}}{\gamma}$
is chosen such that the sound speed inside the computational domain is the same as the 
sound speed $a_s$ at the outer boundary. The initial density and pressure in the 
computational domain are to some extent irrelevant because the injected sub-Keplerian flow 
will wash out this initial condition in several dynamical times. On the 
left radial boundary at ${\rm R_{in}=2 r_g}$, we use outflow boundary condition so as to 
suck matter inside the black hole \citep{hawley2000,hk2001}. It may be possible that such
boundary condition will affect the amount of poloidal magnetic flux accumulation 
near the axis, specially when the magnetic field strength is high. 
However, for most cases, we find that along the ${\rm R_{in}=2 r_g}$ axis, 
the poloidal velocity is greater than the Alfven speed. Therefore, we believe that, 
except for strong magnetic field, the influence of inner boundary on the flow is limited.
Also, we use only the outflow boundary condition on the top and bottom Z-boundaries as the 
injected matter behaves as a hot puffed up disc close to a black hole. No matter is allowed to
enter through the top and bottom Z-boundaries.

\begin{figure*}
\begin{center}
\includegraphics[width=140mm]{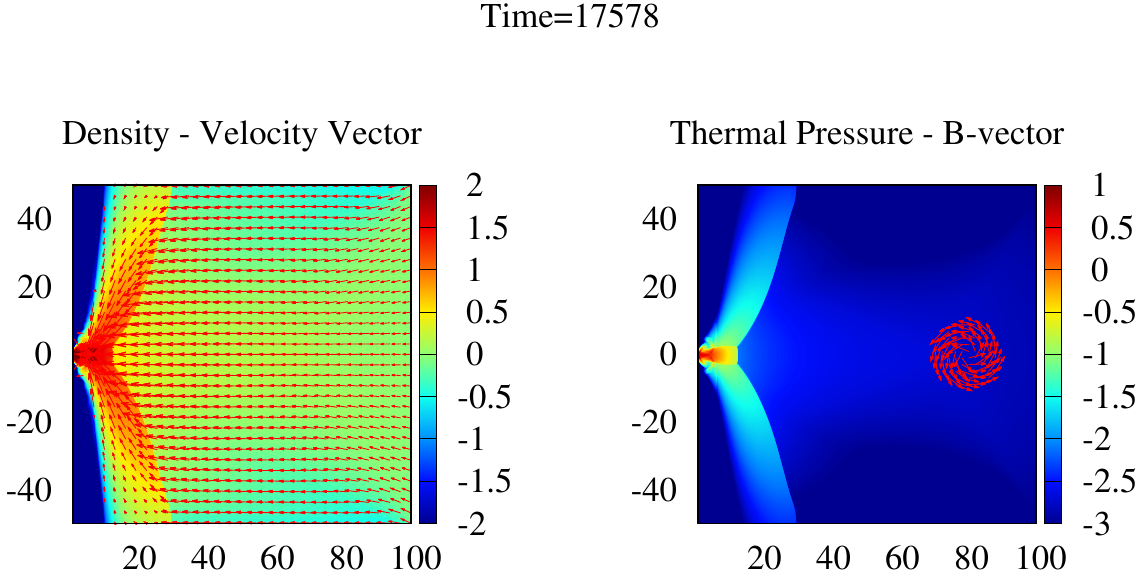}
\end{center}
\caption{shows the animation for run A2 available in the HTML version of this article. 
Left panel shows the density distribution, in log scale, overlayed with velocity vectors. 
Right panel shows the thermal pressure distribution, in log scale, overlayed with 
magnetic field vectors. The animation shows the propagation of four successive flux 
ropes through the sub-Keplerian accretion disk and their effects on the flow dynamics.
The duration of animation is 27 sec.
}
\label{fig0}
\end{figure*}

It is pertinent to ask whether the outflow is affected
by the top and bottom Z-boundaries. Indeed, this would be true 
for any simulation with outflow. However, for this particular sub-Keplerian model,
a very detailed analysis on the boundary conditions have been carried out in section 4 of \citet{ryu1995apj}.
Subsequently, similar boundary conditions have been used in many other works, e.g.,
\citet{mrc1996,rcm1997,giri2010,ggc2012,giri2013,lee2016} etc.
In these works, the Z-boundaries are placed at various heights ranging from 50 r$_g$
to 200 r$_g$. We have used the same Z-boundary boundary conditions as in these works.
Therefore, we believe that the boundary conditions are not affecting the
formation and sustenance of the outflow. Detailed visual inspection of the simulated
images also support the same conclusion. Readers are directed to see the movie in \autoref{fig0} for run A2, which is
available online. (Additionally, it is also available at https://youtu.be/upWm9dFrkt4). 

We use a hydro-steady state as the initial condition for our magnetized accretion 
flow simulations. The hydro-steady state is achieved by running the code for $\sim$10 
dynamical times without magnetic field. The dynamical time is defined as the 
time taken by the matter entering at the outer radial boundary of our computational 
domain to reach the black hole horizon when the flow has achieved a steady state. 
In the results section, we will show the hydro-steady states that develop with an 
accretion shock.

Magnetic fields are very pervasive in astrophysical plasma and it is inevitable that 
the magnetic fields will be dredged in with the accretion flow. We will thus study the 
effects of episodic magnetic flux rope injection on this hydro-steady state that 
develops as a consequence of sub-Keplerian inflow. Specifically, we are interested 
in checking if the outflow rate is enhanced as a result of enhanced magnetic fields in the inflow. 
In this paper, we study the effects of episodic inflow of magnetic flux tubes. 
This hopefully mimics the realistic situation when magnetic fields of various cross-sections and strengths 
are randomly injected continuously into the accreting gas. 
Similar work has been reported in \citet{kudoh2002} where they studied jet formation from
an accretion disk with initial poloidal magnetic field loops. However, our simulation set up is
significantly different compared to this work. In our work, the disk has significant radial
velocity and our flux tubes have an initial toroidal component in addition to
poloidal component.

The magnetic flux ropes, with a helical field, are episodically setup inside the 
computational domain. We use the following magnetic vector potential to set up the 
in-plane components $B_R$ and $B_z$ of the initial magnetic flux ropes.
\begin{displaymath}
A_\phi=\left\{ \begin{array}{ll}
-\sqrt(4\pi)(10-r_{ab}) & {\rm for} ~ r_{ab}\leq10,\\
0 & {\rm for} ~ r_{ab}>10,
\end{array} \right.
\end{displaymath}
where, $r_{ab}=\sqrt{(R-a)^2+(Z-b)^2}$, $a$ and $b$ being the coordinates of the center of the flux ropes.  
In addition to these, a toroidal component $B_\phi \sim \frac{1}{R}$ is added. 
These field components are re-scaled later so that the average plasma beta matches 
with the desired value inside the region surrounded by the flux ropes. We set 
up multiple episodes of magnetic field injection and some of our longer running 
simulations have four episodes where magnetic field is injected. To break the symmetry, 
the flux ropes are randomly injected above or below the midplane of the simulation. 

\section{Results}
\label{sec:sec2}

In Table I, we document the parameters that are used for the simulations presented here. 
The specific energy $\epsilon=0.0021$ is assumed for all the cases. We consider 
two different specific angular momenta $\lambda=1.65$ and $\lambda=1.5$ which are much lower as compared to 
marginally stable value of $1.83$. So, under the normal circumstance, even a zero energy and zero viscosity flow will allow the matter
to fall onto the black hole without any obstruction.
For $\lambda=1.65$, we have $v_{\rm in}=0.048$ and $a_s=0.044$ as the injection parameters for cases A1, A2 
and A3. For $\lambda=1.5$, we have, $v_{\rm in}=0.049$ and $a_s=0.044$ for cases B1, B2 and B3. 
Non-magnetized flow with $\lambda=1.5$ does not show the formation of shock in a flow in vertical equilibrium. 
However, we see that a gas becomes dense due to the centrifugal barrier. 
On the other hand, for $\lambda=1.65$, we see the formation of a steady shock at 13 $r_g$
on the equatorial plane. The post-shock region is sub-sonic, hot and is the location where the thermal 
energy of the inflow is efficiently dissipated in presence of soft (low energy) seed photons. 
This is also the region which supplies matter to outflows. This region is known as the 
CENtrifugal pressure dominated BOundary Layer or CENBOL. 
For each angular momentum mentioned above, we run three MHD simulations
with the magnetic field loops having average plasma betas given by $\beta=50,\ 25,$ and $10$. 
Here, plasma beta is defined as $\beta=P_{gas}/P_{mag}$, where, $P_{gas}$ is the
gas pressure and $P_{mag}$ is the magnetic field pressure.
Therefore, these cases give us an opportunity to study the effects of increased 
magnetic field strength on the accretion disc.

\begin{table}[h]
\begin{tabular}{cccccc}
\multicolumn{6}{c}{Table 1: Parameters used for the simulations.}\\
\hline Case & $\lambda$ & $\beta$ & $v_{\rm in}$ & $a_s$ & $v_\phi=\frac{\lambda}{\rm R_{out}}$\\
\hline
A1 & 1.65 & 50 & 0.04862 & 0.04446 & 0.0165 \\
A2 & 1.65 & 25 & 0.04862 & 0.04446 & 0.0165 \\
A3 & 1.65 & 10 & 0.04862 & 0.04446 & 0.0165 \\
B1 & 1.5 & 50 & 0.04916 & 0.04445 & 0.015 \\
B2 & 1.5 & 25 & 0.04916 & 0.04445 & 0.015 \\
B3 & 1.5 & 10 & 0.04916 & 0.04445 & 0.015 \\
\hline
\end{tabular}
\end{table}

\begin{figure*}
\includegraphics[width=55mm]{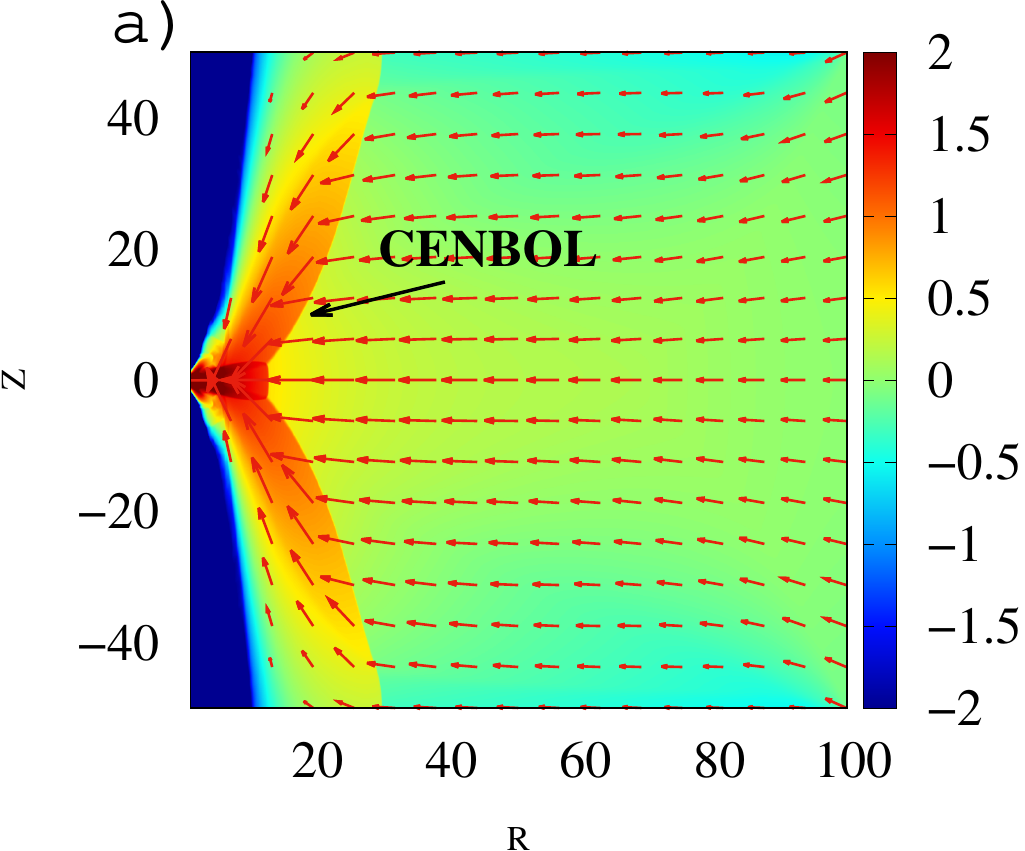}
\includegraphics[width=55mm]{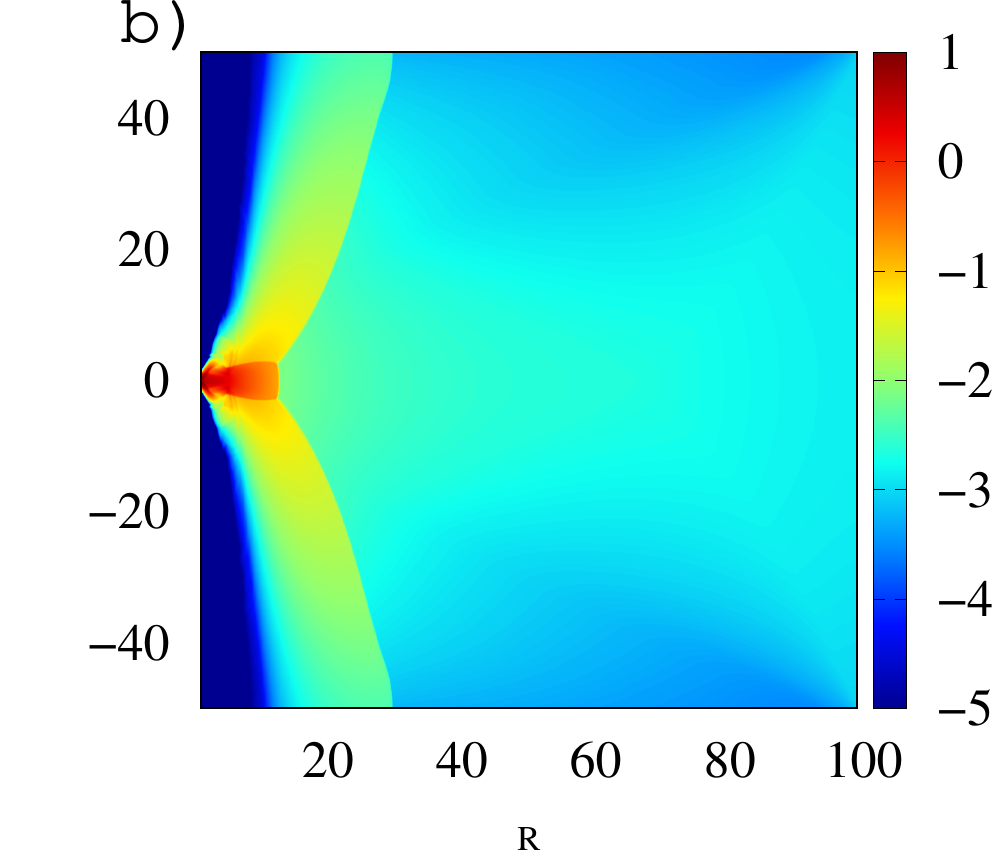}
\includegraphics[width=55mm]{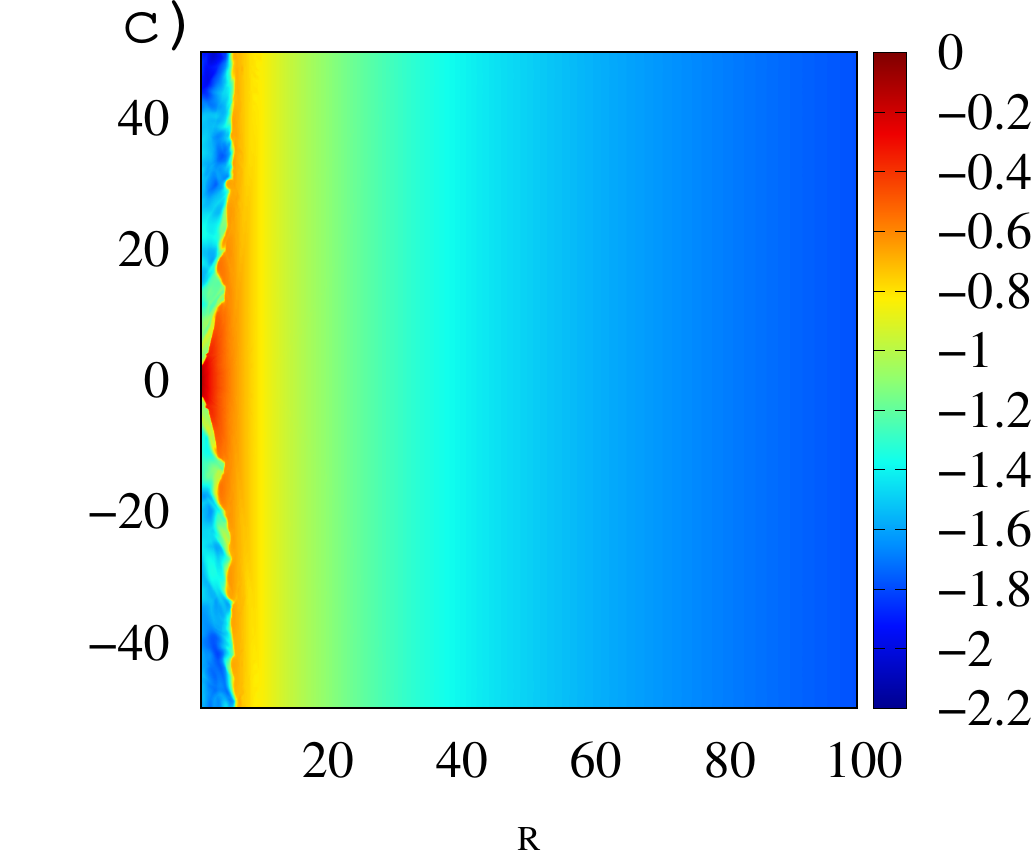}
\caption{Initial hydro-steady state for the simulation runs A1, A2 and A3. 
(a) Density distribution in logarithmic scale overlayed with 
momentum vectors. The length of the vectors is proportional to the
magnitude of the momentum. The Centrifugal pressure supported boundary layer (CENBOL), 
can be seen close to the black hole where there is sudden jump in color in density. 
(b) Pressure distribution in log scale. The CENBOL is found to bend outward 
as one moves away in the vertical direction.
(c) Distribution of the toroidal velocity in log scale inside the disc. 
For our non-viscous simulations, the specific angular momentum remains 
conserved. As the matter approaches the black hole, its rotational velocity increases.}
\label{fig1}
\end{figure*}

\autoref{fig1}(a-c) shows the initial hydro-steady state for the simulation runs A1, A2 and A3. 
\autoref{fig1}a shows the density distribution, in log scale
overlayed with momentum vectors. The length of the vectors is proportional to the
magnitude of the momentum. The centrifugal pressure supported boundary layer (CENBOL), 
can be identified by the jump in color of density. 
The CENBOL is bent outward as one moves away in the vertical direction 
from the equatorial region because the gravitational pull is decreased with increasing 
distance from the black hole. \autoref{fig1}b shows the pressure distribution, again in log scale, 
inside the accretion disc. Here, again, we can identify the shock front (i.e., CENBOL boundary). 
\autoref{fig1}c shows the distribution of the toroidal velocity in log scale inside the disc. 
In our non-viscous simulations, the specific angular momentum remains 
conserved and as a result, as the matter approaches the black hole, its rotational velocity increases.

\begin{figure*}
\includegraphics[width=55mm]{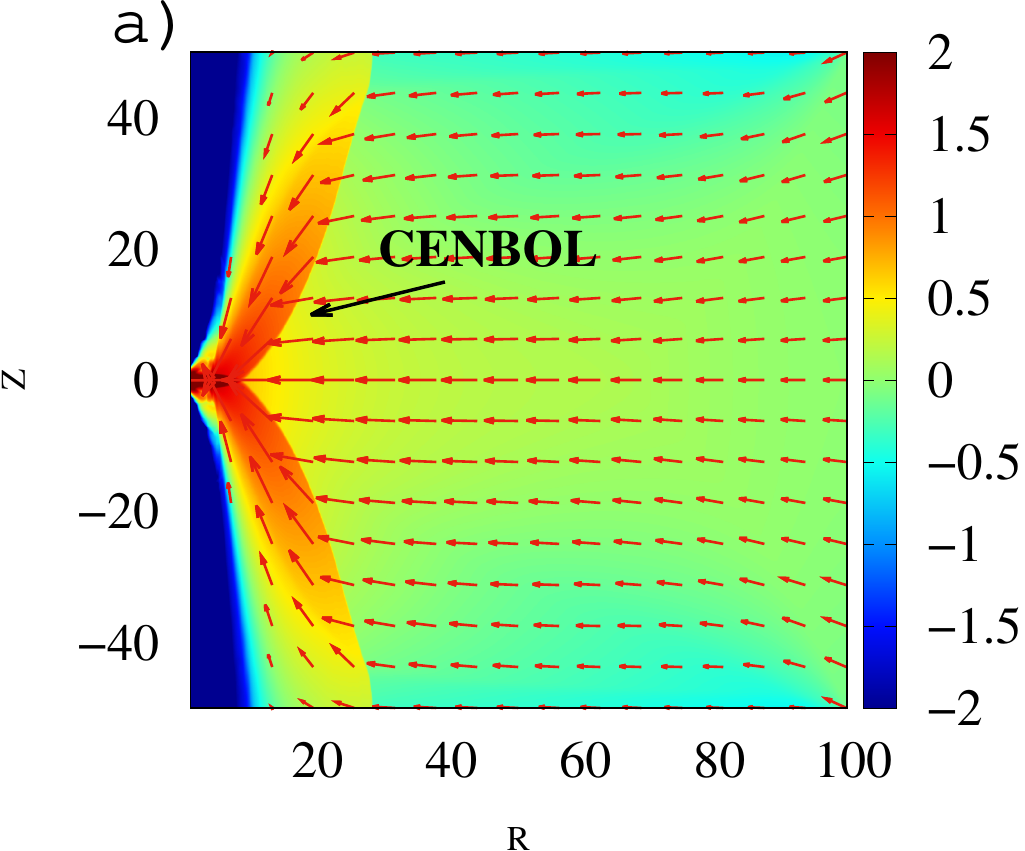}
\includegraphics[width=55mm]{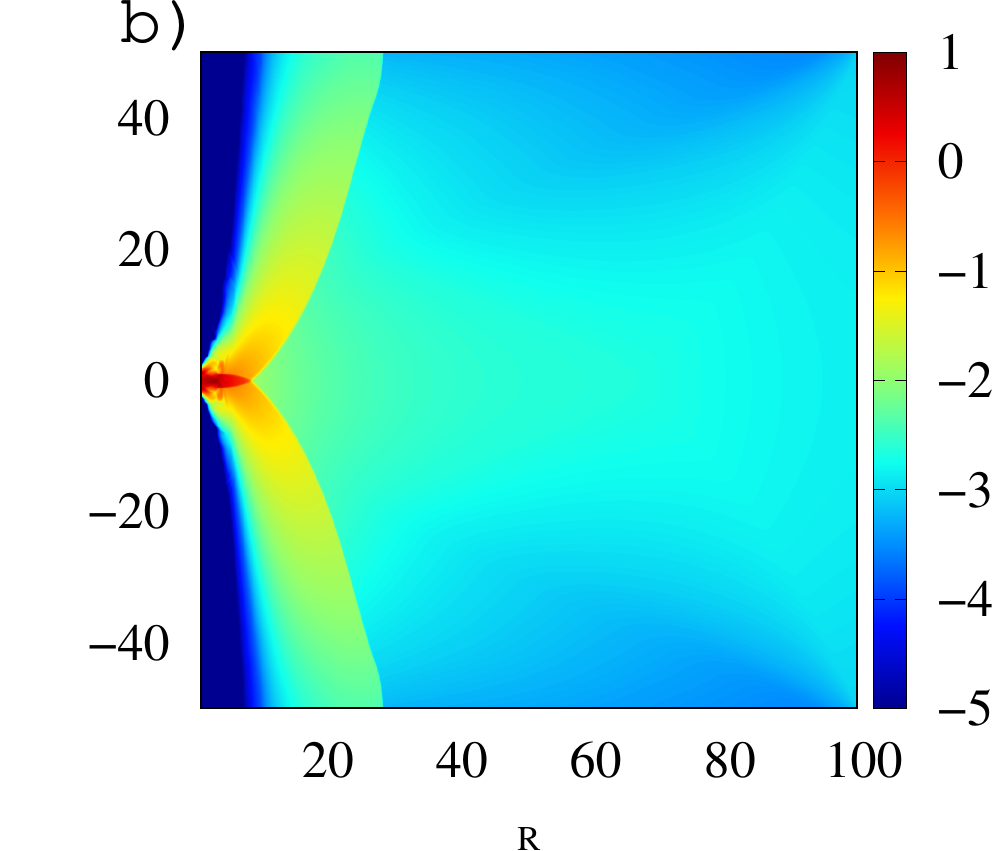}
\includegraphics[width=55mm]{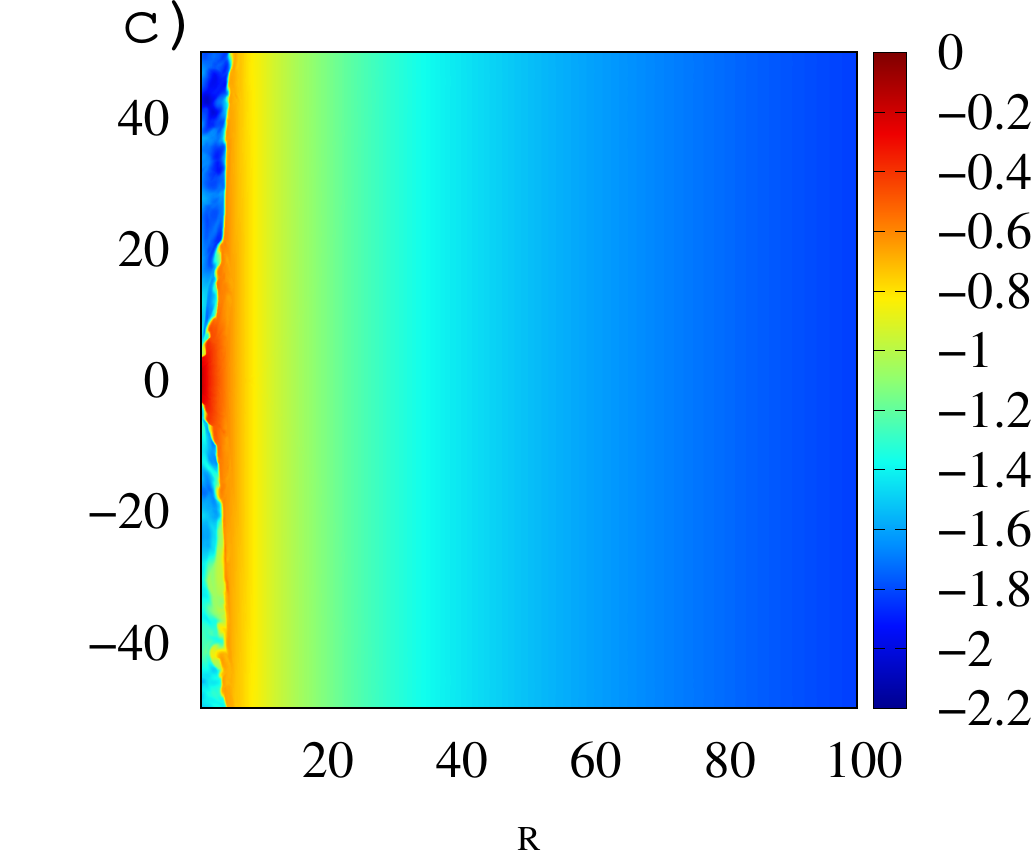}
\caption{Same as \autoref{fig1}, however, for the simulation runs B1, B2 and B3. 
}
\label{fig2}
\end{figure*}

\autoref{fig2} shows the initial hydro-steady state for the simulation runs B1, B2 and B3. 
As in \autoref{fig1}a, here also we can identify the CENBOL in \autoref{fig2}a. Note that 
the parameters used for this simulation does not allow shock formation theoretically 
when the vertical equilibrium model is used. 
However, we find that matter indeed slows down because of the centrifugal force forming
the CENBOL. \autoref{fig2}b shows the pressure distribution in the log scale 
inside the accretion disc and \autoref{fig2}c shows the distribution of the toroidal 
velocity, again in a log scale, inside the disc.

\subsection{Propagation of flux ropes through sub-Keplerian flow}

For cases A1, A2, B1 and B2, we have four episodes of flux rope injection at four 
successive times. For case A1, the flux ropes are injected at times 17578, 26930,
36529 and 46388. For case A2, the flux ropes are injected at times 17578, 32965, 45605
and 57961. For case B1, the flux ropes are injected at times 22227, 33312, 46372 and 59242.
For case B2, the flux ropes are injected at times 22227, 37581, 53079 and 67045. 
For all these cases, the centers of the flux ropes are placed at 
four different locations given by the $\left[ {\rm R, Z}\right]$ coordinates as
$\left[ 80,0\right]$,$\left[ 85,20\right]$,$\left[ 85,-22\right]$,$\left[ 80,0\right]$
inside the accretion disc. For case A3, we have one injection at time 17578 and the center is 
placed on the equatorial plane at $\left[ 80,0\right]$. For case B3, we have two 
injections at two successive times at 22227 and 33254, and the centers are placed at
$\left[ 80,0\right]$ and $\left[ 85,20\right]$, respectively. Because of the stronger magnetic 
fields in cases A3 and B3, they run with incredibly small timesteps, especially 
when the magnetic field reaches the CENBOL region. For this reason, we were not 
able to explore many episodes of field injection in those two runs. The radius of 
the outer rope is 10 $r_g$ at injection in all the cases. The flux ropes are initialized 
so that they are in pressure equilibrium with the surrounding material. Thus, 
the matter density in the region surrounded by the magnetic field lines are reduced 
in such a way that the sum of the thermal and magnetic pressure is the same as the 
thermal pressure before initialization of magnetic flux ropes. This ensures pressure 
balance with surrounding material inside the sub-Keplerian disc. The accretion flow 
carries the flux ropes toward the central object. Since the accretion flow is 
mainly in radial direction, our experience has been that it does not matter too much whether 
the magnetic flux ropes are injected above or below the mid-plane.

\begin{figure*}
\includegraphics[width=55mm]{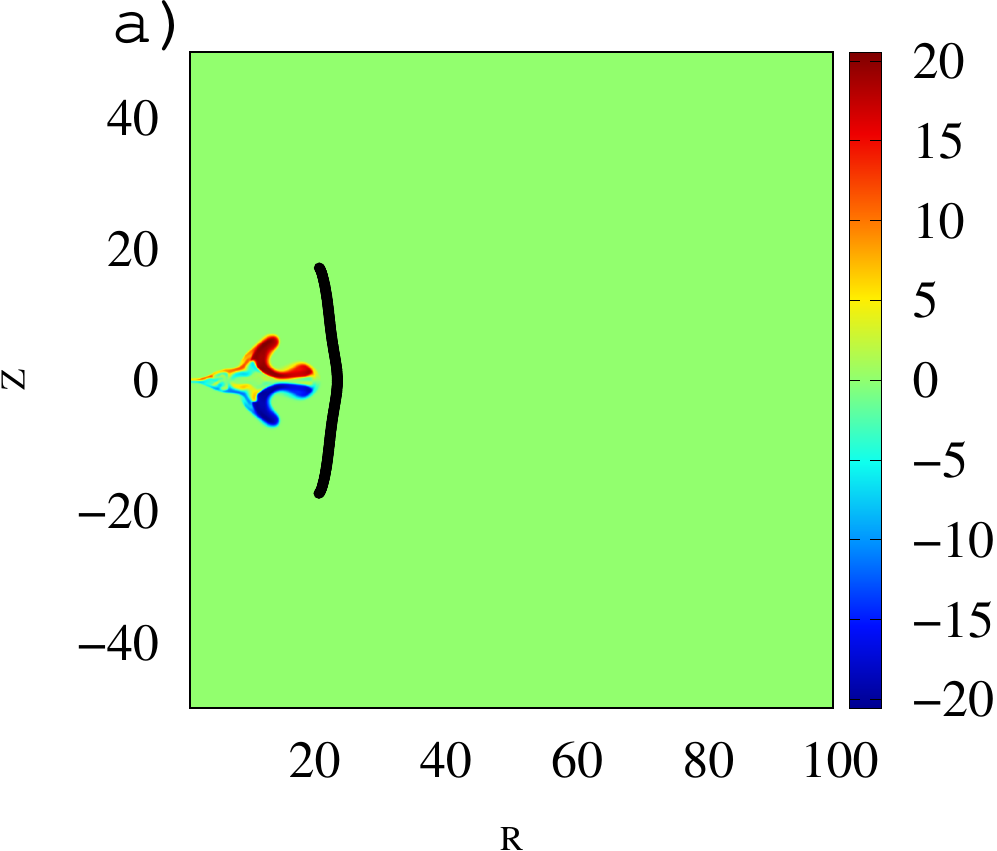}
\includegraphics[width=55mm]{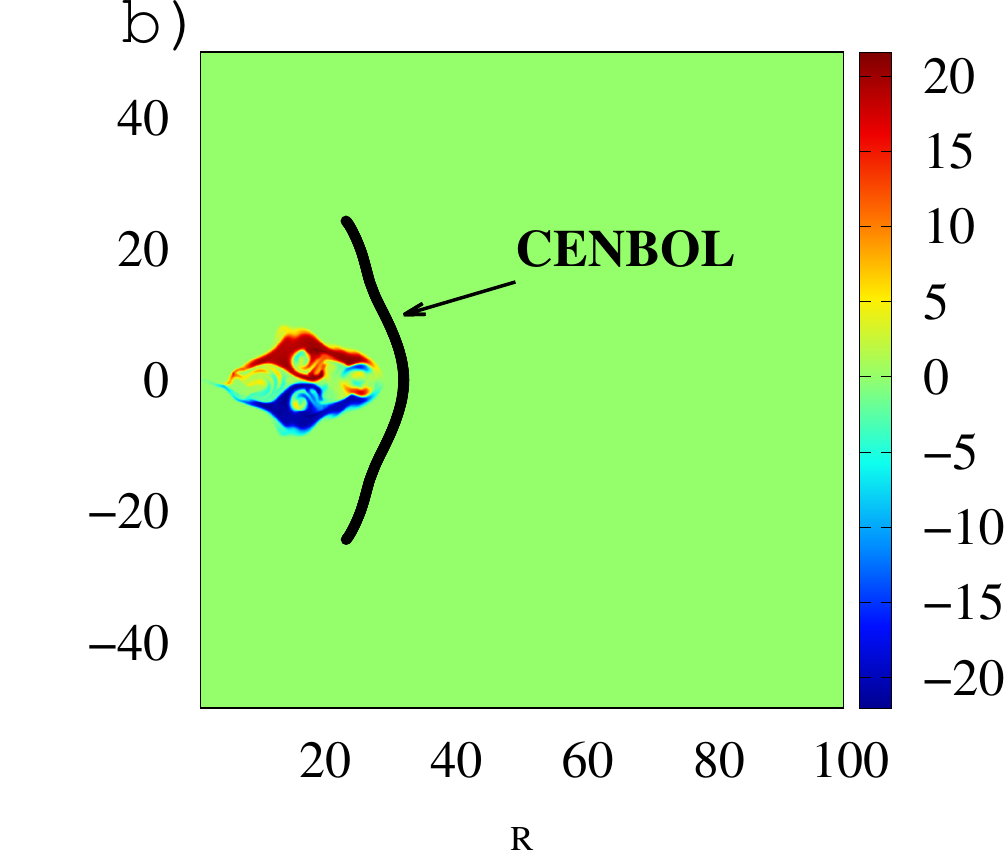}
\includegraphics[width=55mm]{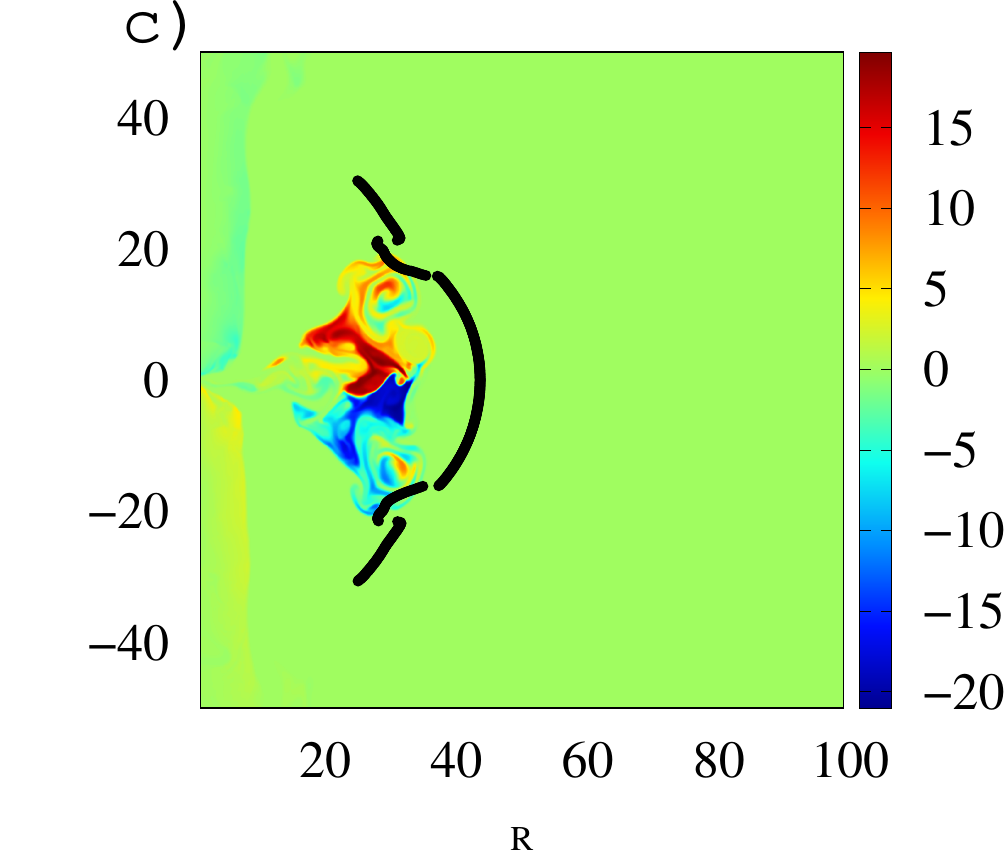}
\caption{Distribution of toroidal component of magnetic fields inside 
the accretion disc for three different cases, A1, A2 and A3, with plasma beta of (a) 50, (b) 25 and (c) 10, respectively. 
The snapshots are drawn at a time of 18578 $r_g/c$ and the specific angular momentum is $1.65$ 
for all the three cases. The thick, solid, black line shows
the shock surface (i.e., CENBOL).
We see that the toroidal field has opposite signs above and below the 
equatorial plane. This is caused by the differential toroidal velocity. In the toroidal 
direction, the stretching of the field vary with radii because of the 
differential toroidal velocity. However, since the magnetic field directions above and 
below the equatorial planes are opposite, the toroidal field also attains opposite sign. 
}
\label{fig3}
\end{figure*}

As the flux ropes radially propagate through the sub-Keplerian disc, we notice that the flux 
ropes are stretched in the radial direction. The front end moves with the higher 
radial speed whereas the rear end moves with the lower radial speed, thus 
stretching the flux ropes in the radial direction. We further notice that the 
toroidal component of the flux ropes attains a positive value above the mid-plane 
whereas the same attains negative value below the mid-plane. This happens due to 
the differential rotation of matter inside the disc. Matter close to the black 
hole has a higher toroidal velocity as compared to the matter that is at higher radial 
distance from the black hole. Thus, the matter attached to the front end of the 
flux ropes stretches the field lines in the toroidal direction more compared to 
same on the rear end. Since the field lines of the initial flux ropes above the 
mid-plane are directed towards the black hole ($B_R$ has negative sign), this 
differential stretching generates a positive toroidal magnetic field component 
Similarly, since the field lines below the mid-plane are directed away from 
the black hole, the differential stretching generates a negative toroidal 
magnetic field component. \autoref{fig3}(a-c) show the toroidal field component 
for cases A1, A2 and A3, respectively, at a time of 18578 $r_g/c$. 
By this time, the first episode of flux ropes has reached the CENBOL region. 
The amplitude the toroidal component is further increased due to shock compression 
in this region. We clearly see that the toroidal component above the equatorial 
plane has positive values, whereas the same below the equatorial plane has negative values.
Note also that as the field strength is increased (from a to c) the flux tube becomes
more buoyant and expands away from the equatorial plane.

\begin{figure*}
\includegraphics[width=55mm]{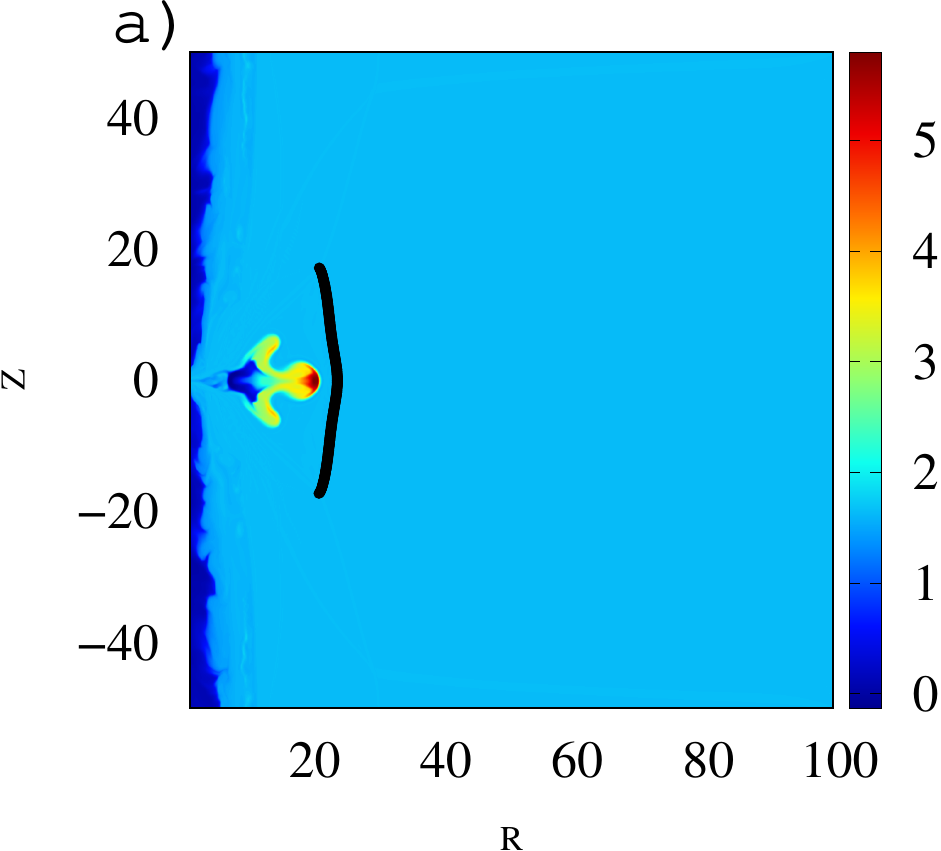}
\includegraphics[width=55mm]{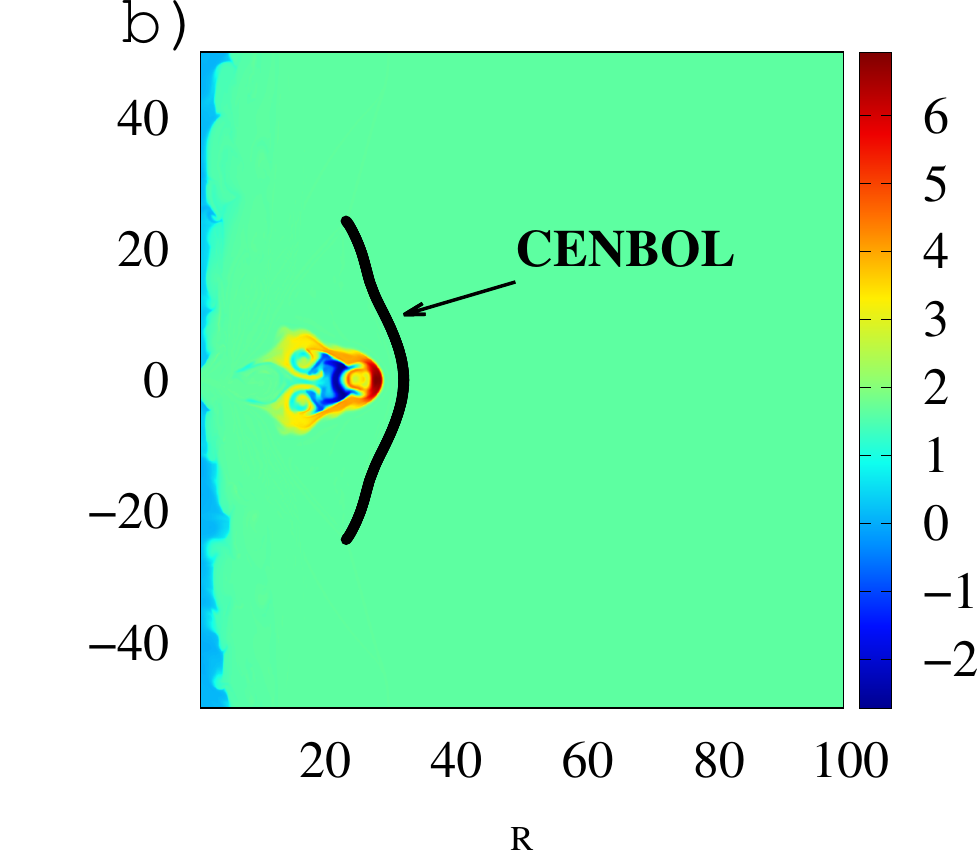}
\includegraphics[width=55mm]{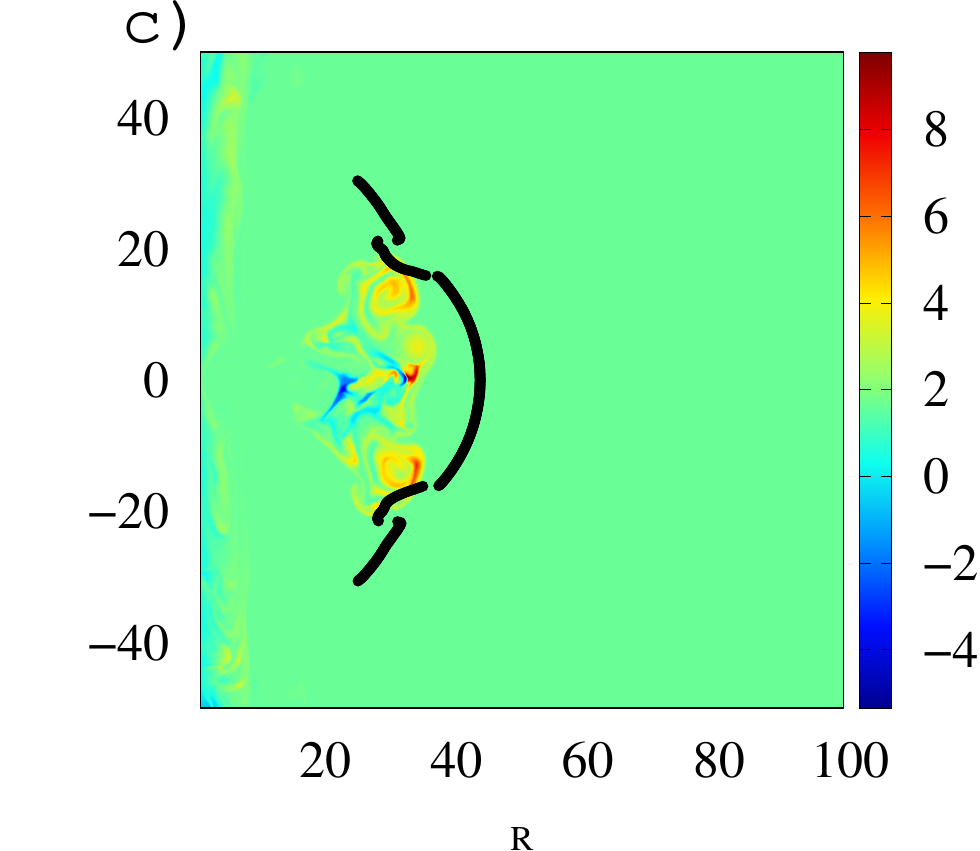}
\caption{Specific angular momentum distribution inside the accretion 
disc for cases A1, A2 and A3, with plasma beta of (a) 50, (b) 25 and (c) 10. 
The snapshots are at $t=$ 18578 $r_g/c$. We find
that the angular momenta mostly have a constant value of 1.65 everywhere except 
in some locations in the post-shock region. The thick, solid, black line shows 
the shock surface (i.e., CENBOL). Specifically, we can see the generation 
of negative and large positive angular momentum. 
As the strength of the magnetic field is increased (e.g., plasma beta of 10), 
the magnitudes of the positive and negative angular momentum are also increased. 
Notice the change of color scale from (a) to (c). See text for details.
}
\label{fig4}
\end{figure*}

An interesting finding emerging from our simulations is the presence of matter with
negative specific angular momentum of the flow as the flux ropes propagate through the 
sub-Keplarian disc. Since the net angular momentum should be constant in
a non-dissipative system, the flow may acquire negative angular momentum
at the cost of magnetic angular momentum.
\autoref{fig4}(a-c) show the specific angular momentum 
distribution inside the accretion disc for cases A1, A2 and A3, respectively. 
The snapshots are taken at a time of 18578 $r_g/c$ for all three cases. 
We see that the angular momenta mostly have a constant value of 1.65 everywhere 
except in some locations in the post-shock region, wherever the field has penetrated. 
The thick, solid, black line shows the shock surface (i.e., CENBOL).
Specifically, we can see the generation of negative and large positive angular momentum
inside the CENBOL. This is caused by the component of Lorentz force which acts in the toroidal 
direction in the R-Z plane. When the field loops reach the post-shock region, 
they are compressed and the magnitude of the magnetic field is increased. 
The torque associated with the azimuthal component of Lorentz force
(L$_\phi$) causes the drift of angular momentum from the inner to the outer edge of the field loop.
Within the field loop, the azimuthal component of Lorentz force can be written as 
L$_\phi=v_ZB_R-v_RB_Z$. For our field loops, close 
to the mid-plane, the field lines are mostly vertical ($B_R\sim0$) and therefore L$_\phi\sim-v_RB_Z$. 
Since the matter mostly move towards the black hole, $v_R$ has a negative sign.
Also, the front end of the flux ropes has field lines with negative values of 
$B_Z$ whereas the rear ends have positive values. Therefore, Lorentz force acts 
oppositely on the two ends of the field loop and forces the matter on the front end 
to have negative angular momentum and the matter on the rear end to have large 
positive angular momentum. As the strength of the magnetic field 
is increased (e.g., plasma beta of 10, i.e., case A3), the magnitudes of the positive 
and negative angular momentum are also increased (notice the changing color scale from 
\autoref{fig4}a to \autoref{fig4}c). If we concentrate on the equatorial place, in all the three cases,
we find the angular momentum gradient is positive, i.e., angular momentum is transported 
outward. This is thus a demonstration that a magnetic viscosity can work inside 
an accretion disc. However, the appearance of oppositely rotating matter needs to be 
revisited in more detailed three dimensional simulations in our future work.
Our findings for the propagation of flux ropes are very much similar to what is described
in section 3.1 of \citet{rbl1996}. As discussed in this reference, many of these effects are
manifestation of magnetorotational instability as discussed by \citet{hawley1991,balbus1991}.

\subsection{Effects of magnetic field on the CENBOL}

As the flux ropes arrive at the CENBOL region, the magnetic field is amplified due to the
shock compression. This, in turn, increases the magnetic pressure 
inside the CENBOL region. Thus, the CENBOL is expanded in both radial and vertical 
directions. Some of the infalling matter, along with magnetic field, leave 
the computational domain through the top and bottom Z-boundaries as an outflow/jet. 
This releases the total pressure in the CENBOL region and the CENBOL tries to go back 
to its original, un-magnetized configuration. Since the magnetic field is torn apart, 
it does not leave completely in one go. Rather, it leaves part by part, causing 
oscillations of the CENBOL. It is interesting to note that we never 
find a complete destruction of CENBOL for any of the cases presented here. This is 
mainly because of sustained supply of spiraling flow which creates a permanent centrifugal barrier. This
also demonstrates the stability of the axisymmetric shock in presence of strong magnetic fields.

\begin{figure*}
\includegraphics[width=55mm]{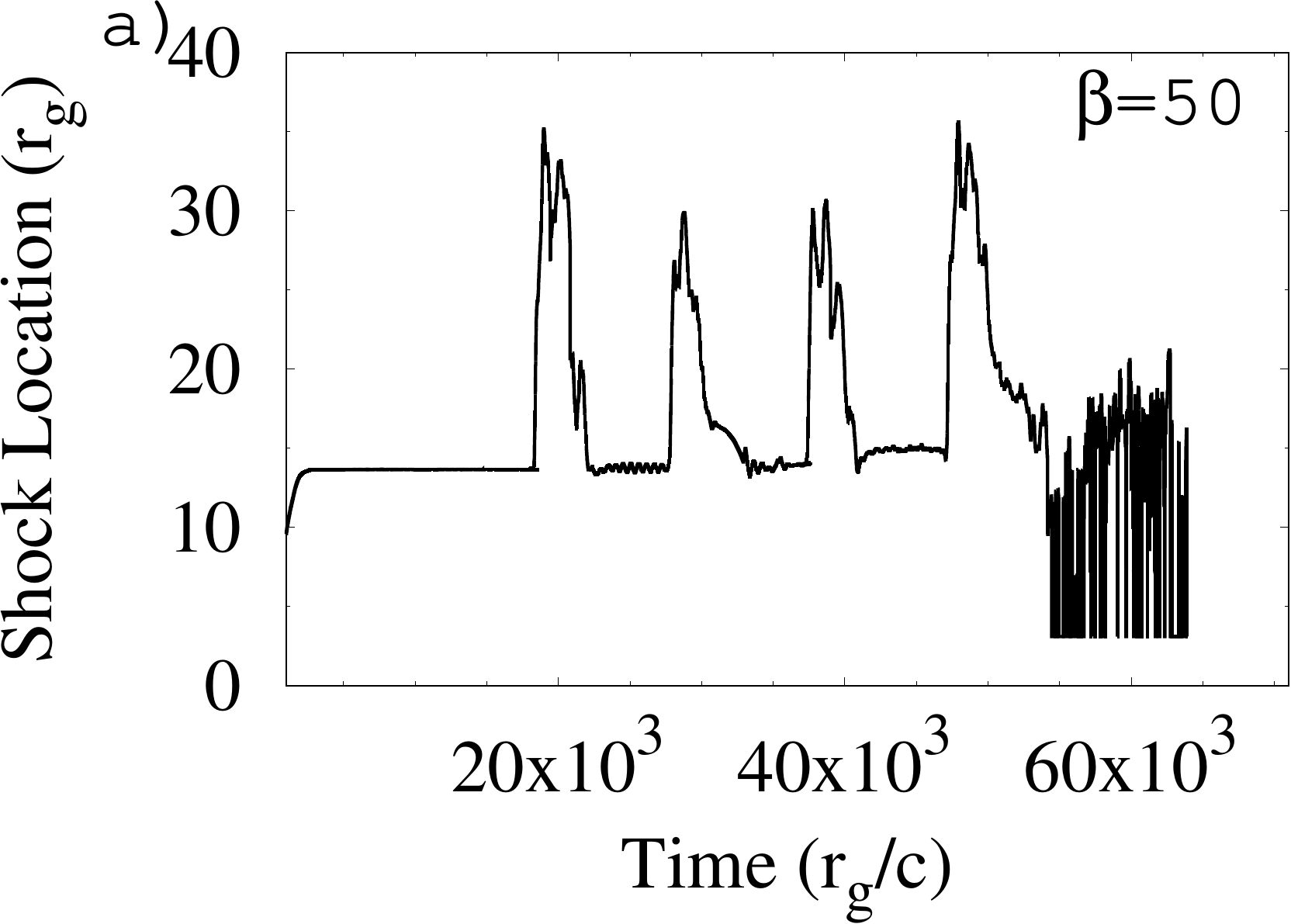}
\includegraphics[width=55mm]{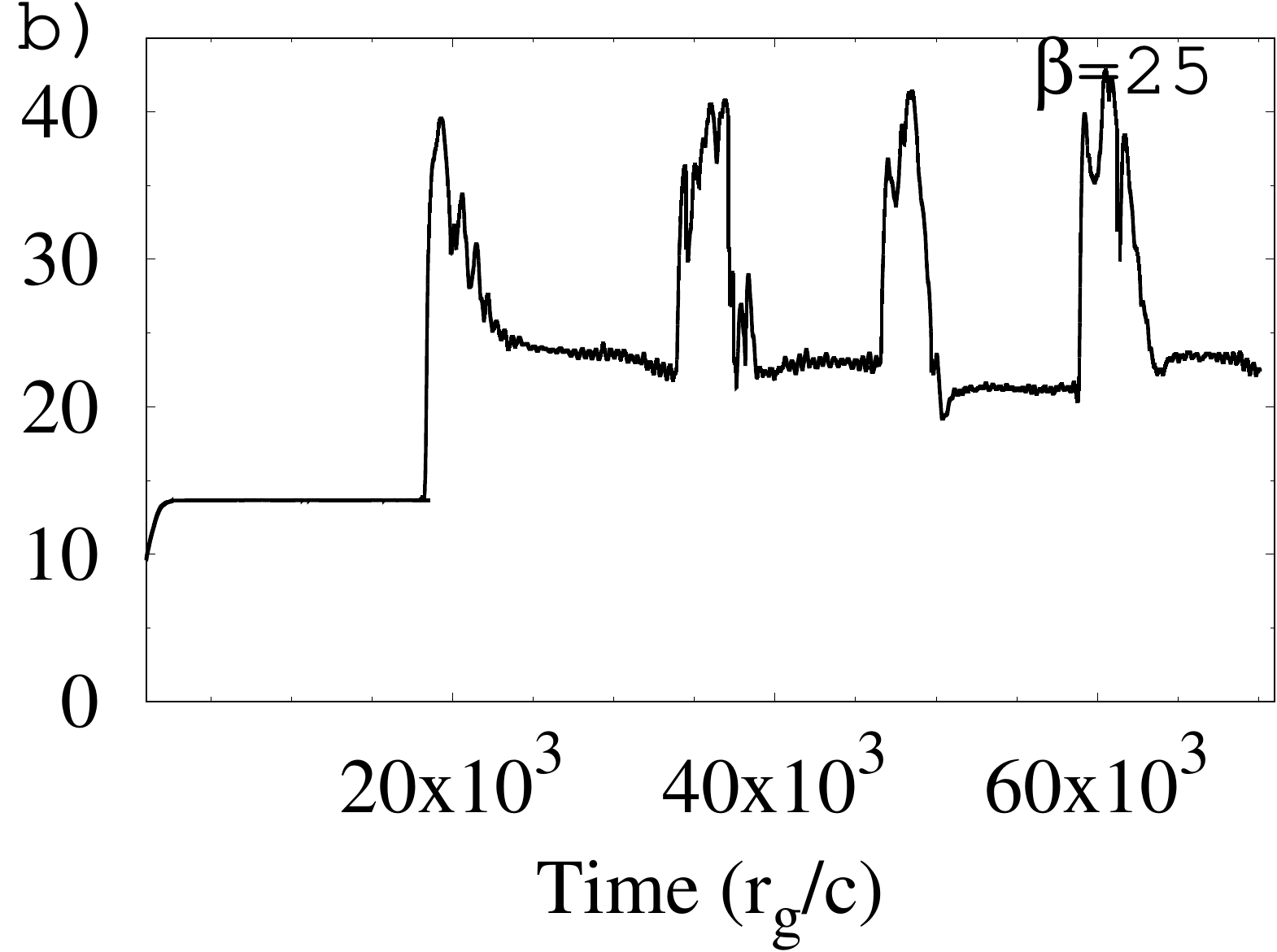}
\includegraphics[width=55mm]{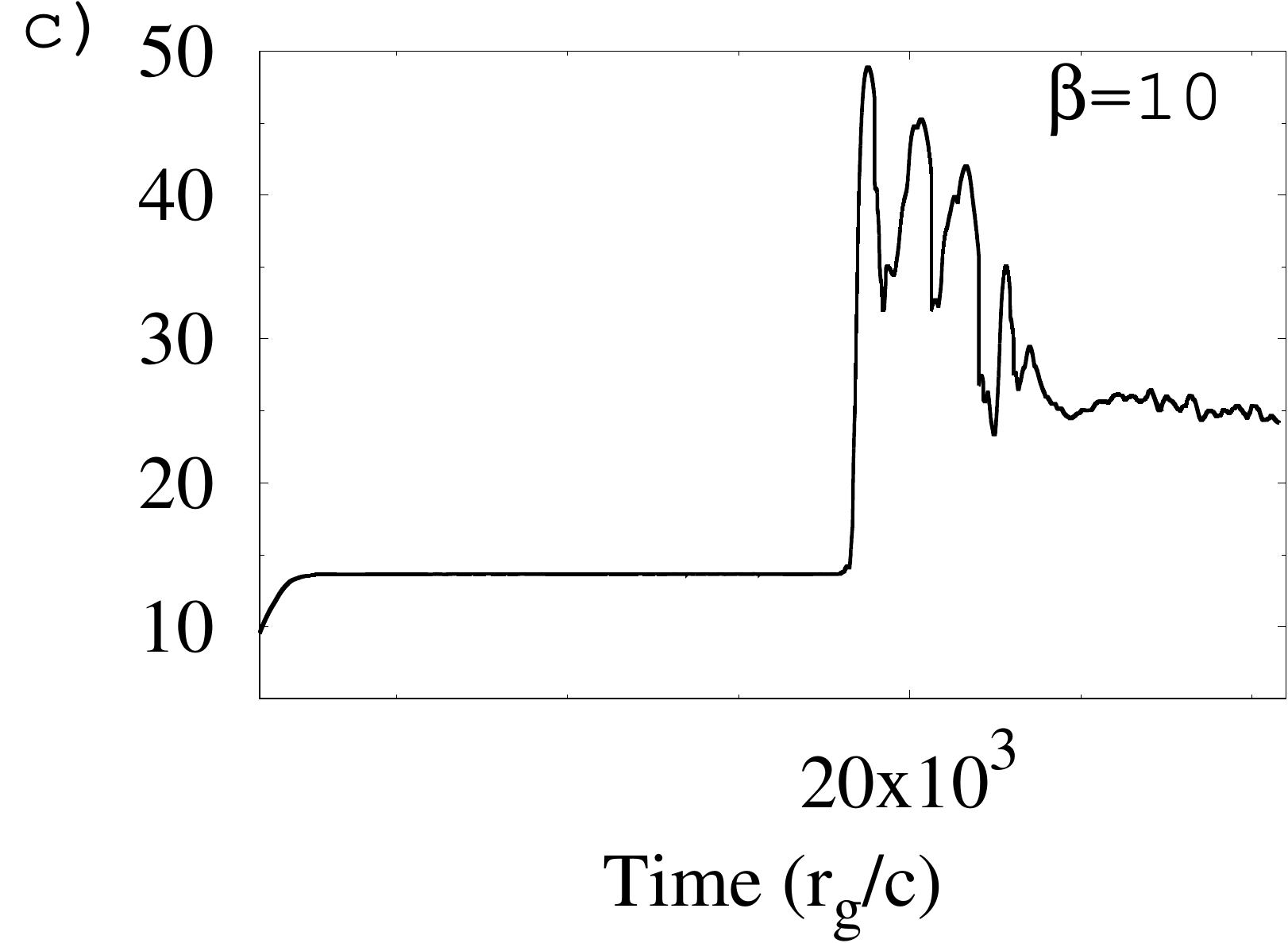}
\caption{Time variation of the shock location on the equatorial plane the accretion 
disc when magnetic flux ropes, with a helical field, are injected into the flow. 
We show three different cases A1, A2 and A3 with plasma beta of (a) 50, (b) 25 and (c) 10. 
The specific angular momentum of 1.65 is used for all the cases shown here. 
(a) and (b) show four episodes of flux rope injection; while (c) shows only 
one such episode. We see that the shock location moves further out with increasing 
magnetic field strength (i.e. smaller plasma beta) due to joint combination of 
the thermal and magnetic pressures.
}
\label{fig5}
\end{figure*}

\autoref{fig5}(a-c) show time variation of the shock location on the equatorial 
plane of the accretion disc for cases A1, A2 and A3, respectively. Initially, we find 
the steady state shock to be located at 13 $r_g$ in all these cases. 
When the first flux rope arrives the CENBOL region, the shock 
location moves momentarily further out from the central black hole. However, it slowly 
moves closer to the black hole and stabilize at somewhat intermediate location before 
the next set of flux ropes arrive at the post-shock region. The intermediate location 
depends on the field strength. If the field is weak and is removed from the region, the 
resulting shock is located at the same place as the hydrodynamic shock. If the field is stronger
(b and c), the shock settles to a larger distance due to leftover field pressure. 
The shock location after successive injection depends on the  strength of the residual 
field in the post-shock region. Furthermore, the shock does not settle to the 
resulting location smoothly. Rather it oscillates a few times, the frequency  increases 
with decreasing field. For weak fields, the oscillations
are not well defined, but they look more prominent as the field becomes stronger, since the magnetic tension 
prevents from quick expansion of the flux tubes. Our work is the first to show that magnetized flow 
also has a quasi-stable shock waves around the black hole. 

\begin{figure*}
\includegraphics[width=55mm]{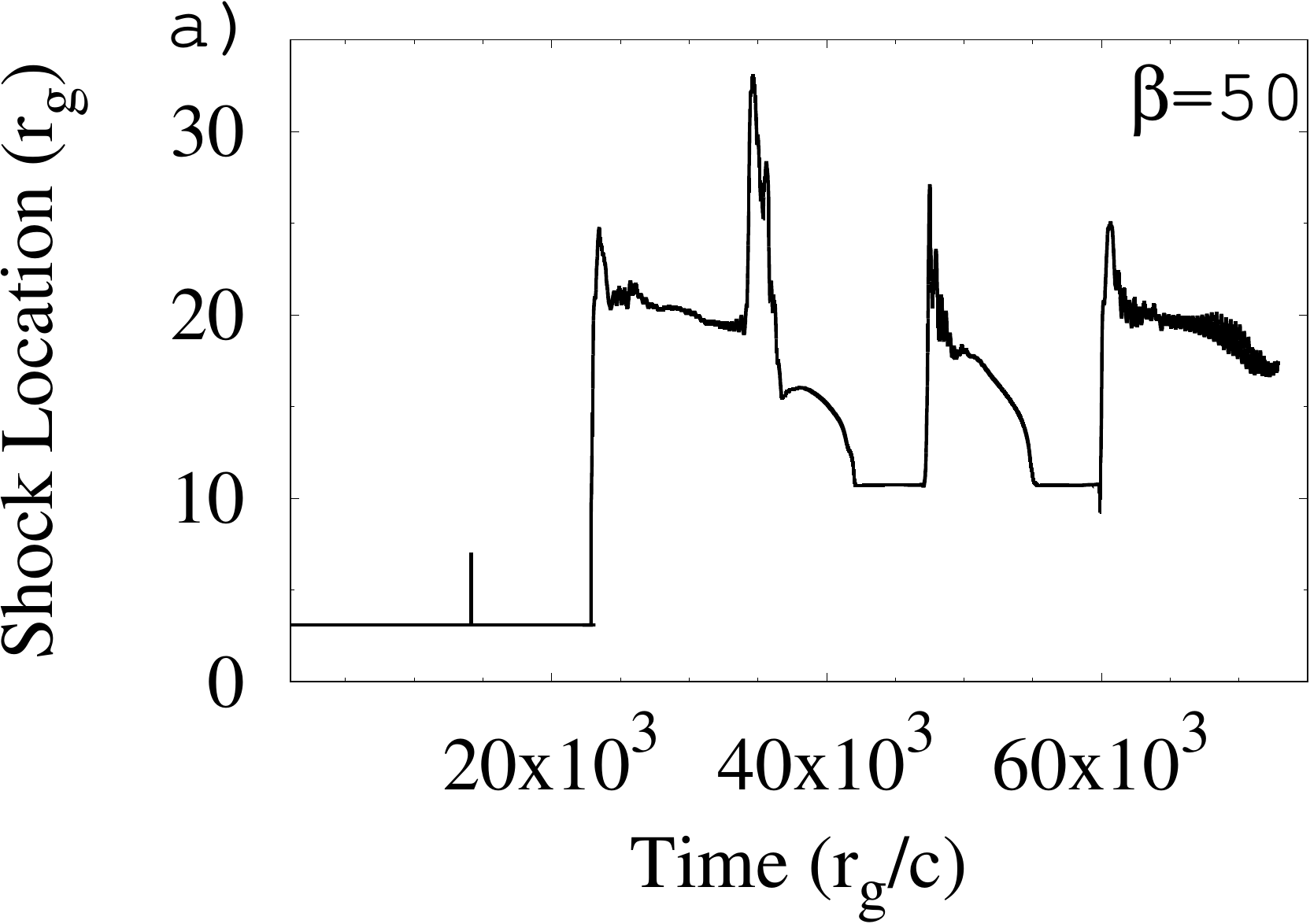}
\includegraphics[width=55mm]{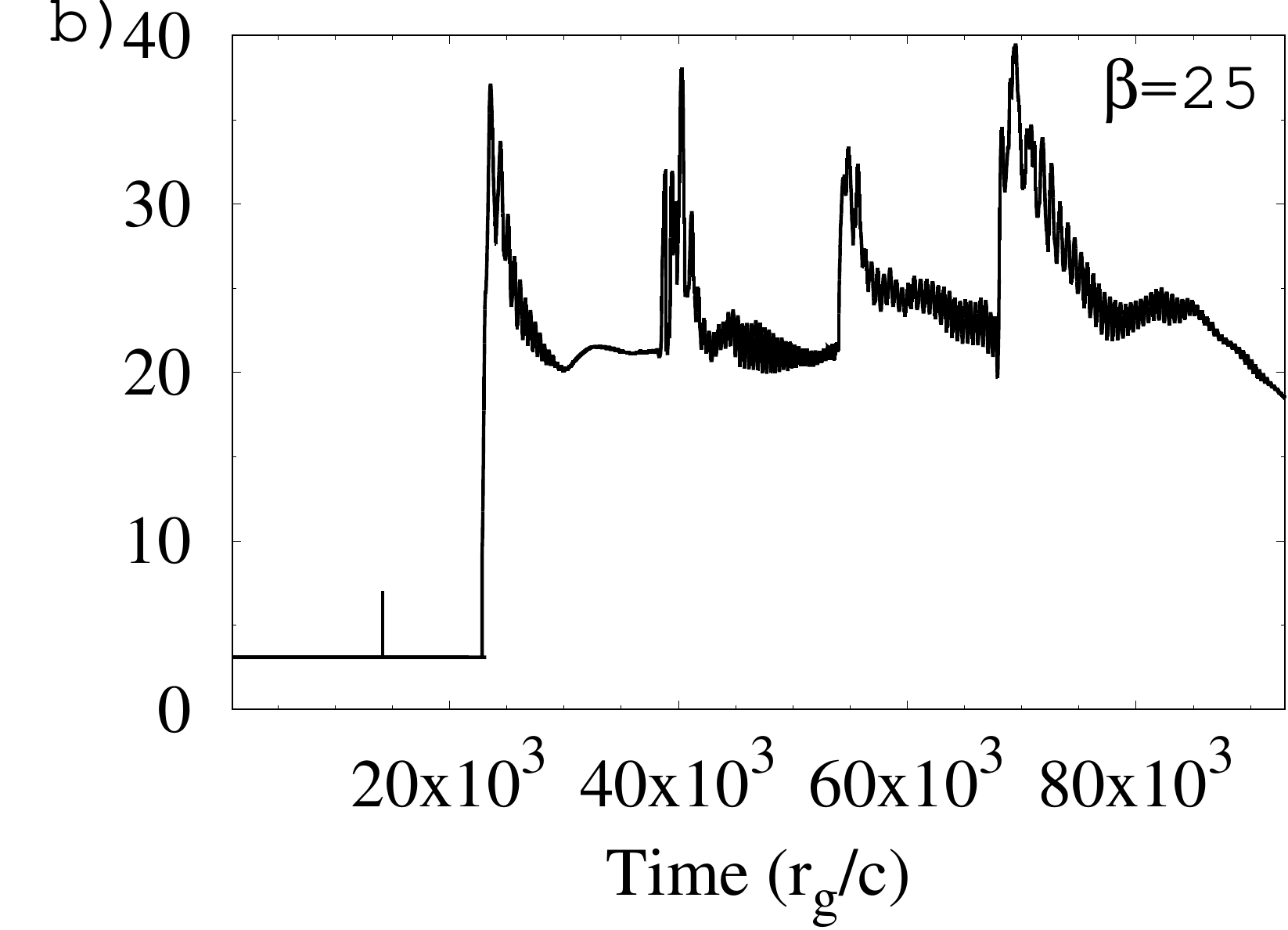}
\includegraphics[width=55mm]{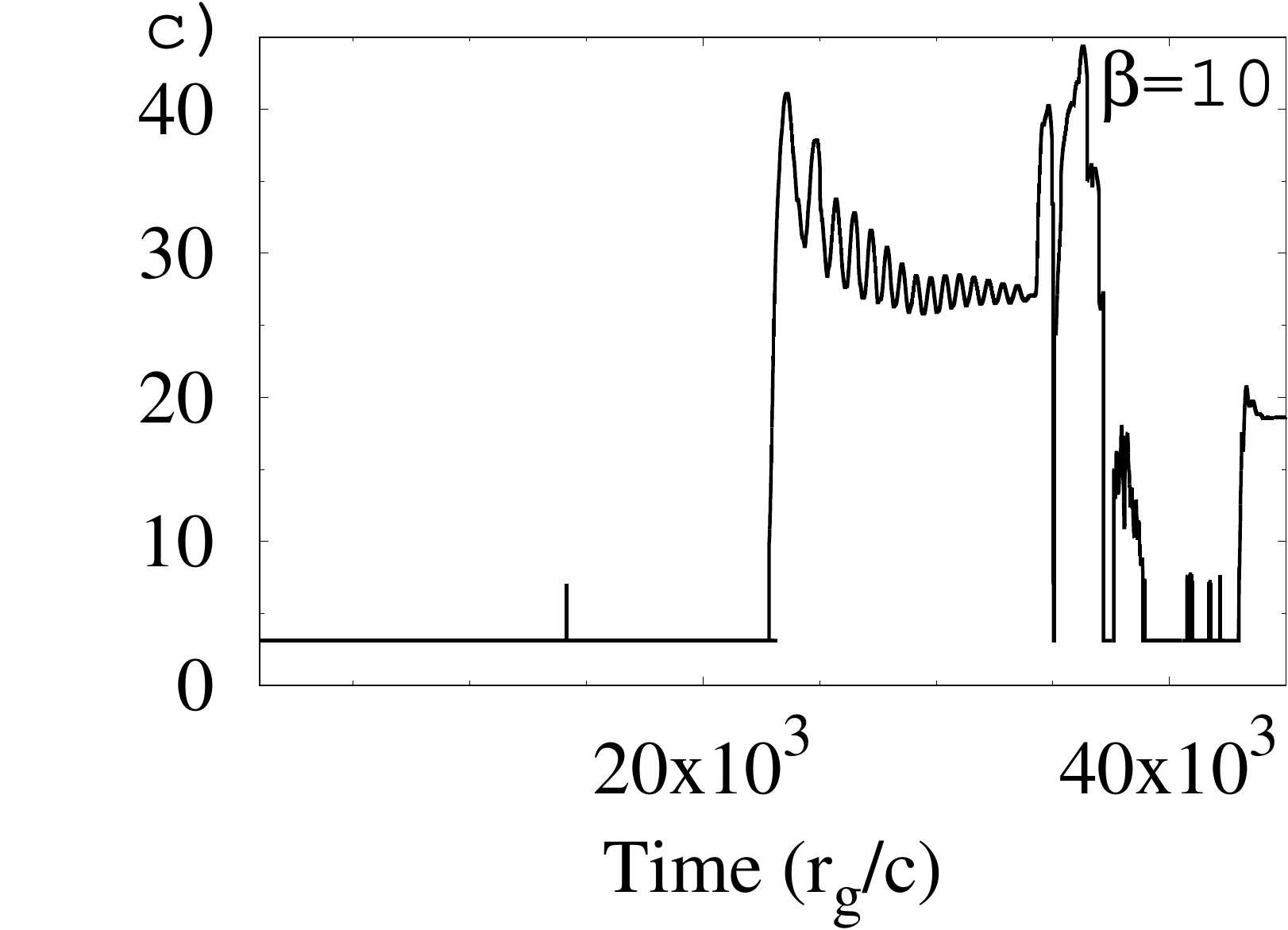}
\caption{ Same as in \autoref{fig5}, but with a specific angular momentum of 1.50 (cases B1, B2 and B3). 
(a) and (b) show four episodes of flux rope injection while (c) shows 
two such episode. In this case, the non-magnetized flow 
does not have a shock on the equatorial plane. However, a shock is formed when the field 
rope is present and it moves further out with increasing magnetic field strength (i.e., smaller plasma beta). 
}
\label{fig6}
\end{figure*}

\autoref{fig6}(a-c) show time variation of the shock location on the equatorial plane 
for cases B1, B2 and B3, respectively. For these cases, the 
hydrodynamic flow does not have a shock. However, due to the presence of non-zero 
angular momentum, the flow has a centrifugal barrier and the matter slows down 
as it approaches the black hole. Thus, a CENBOL is formed for this case as well. 
Since the important ingredient of the shock is the 
centrifugal force, which is very weak in this case, even the 
weakest field pressure was a significant addition of the pressure in the CENBOL which
causes the formation of shocks. As in the previous cases, the shock momentarily moves far away from the black 
hole and then stabilizes at a somewhat intermediate location. In all the cases the shock 
remains due to magnetic pressure. Furthermore, the amplitude of oscillation of the shock 
is low, since the centrifugal force itself is small. For weakest field the oscillation
frequency is higher, but the amplitude is lower. As before, the shock location 
moves further out as the strength of the magnetic field is increased.

\subsection{Effects of magnetic field on the outflow}

\begin{figure*}
\includegraphics[width=55mm]{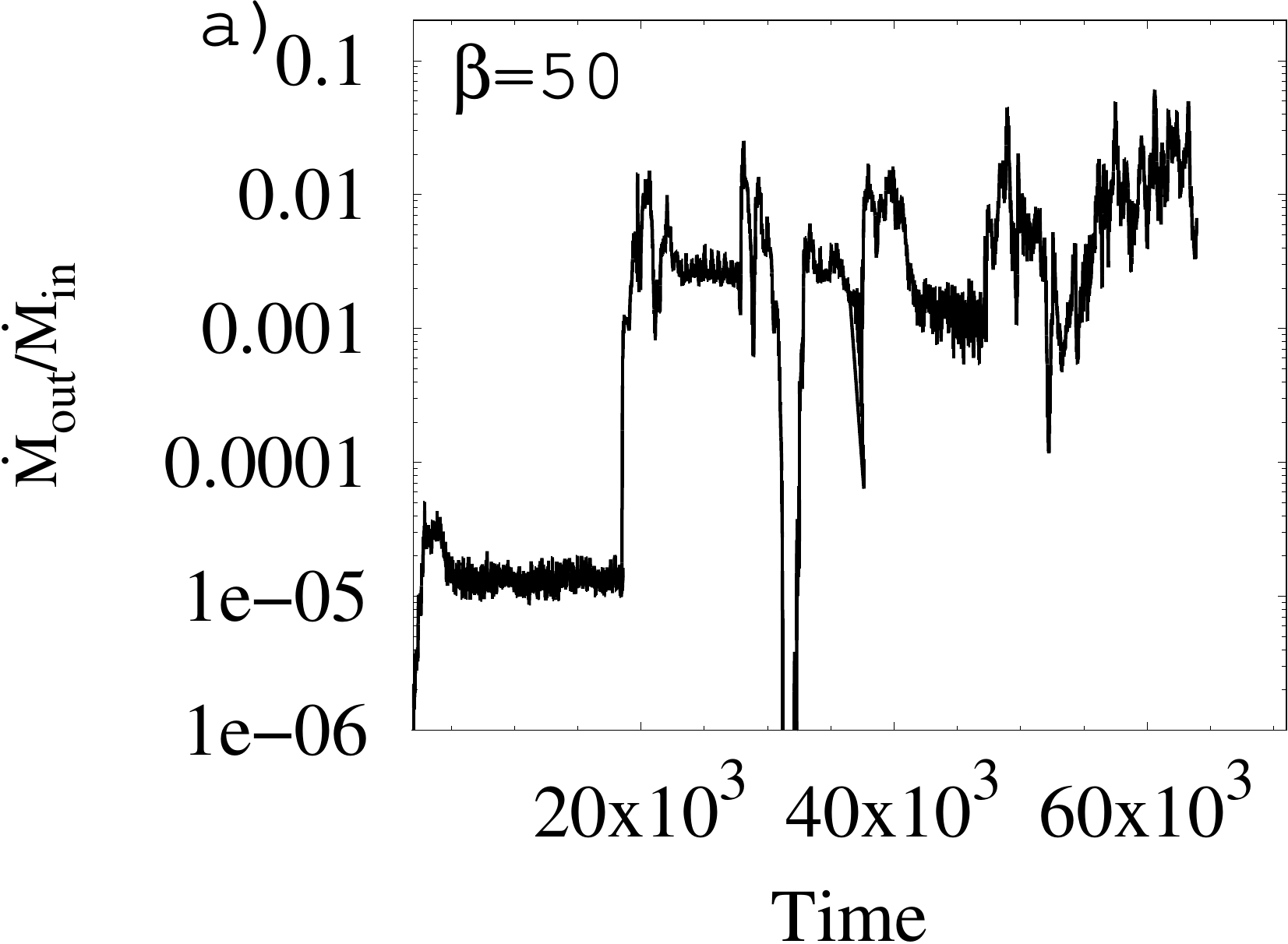}
\includegraphics[width=55mm]{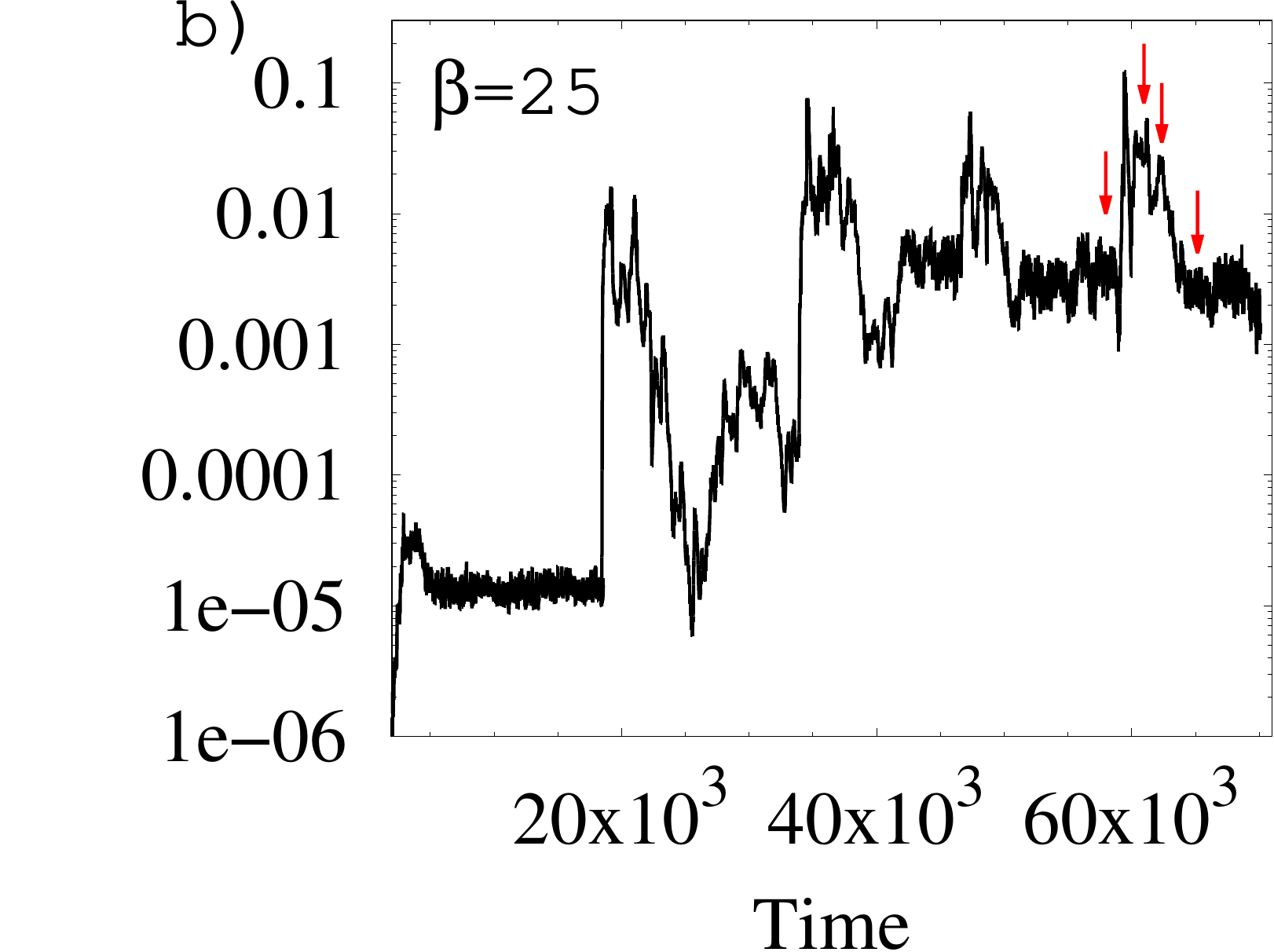}
\includegraphics[width=55mm]{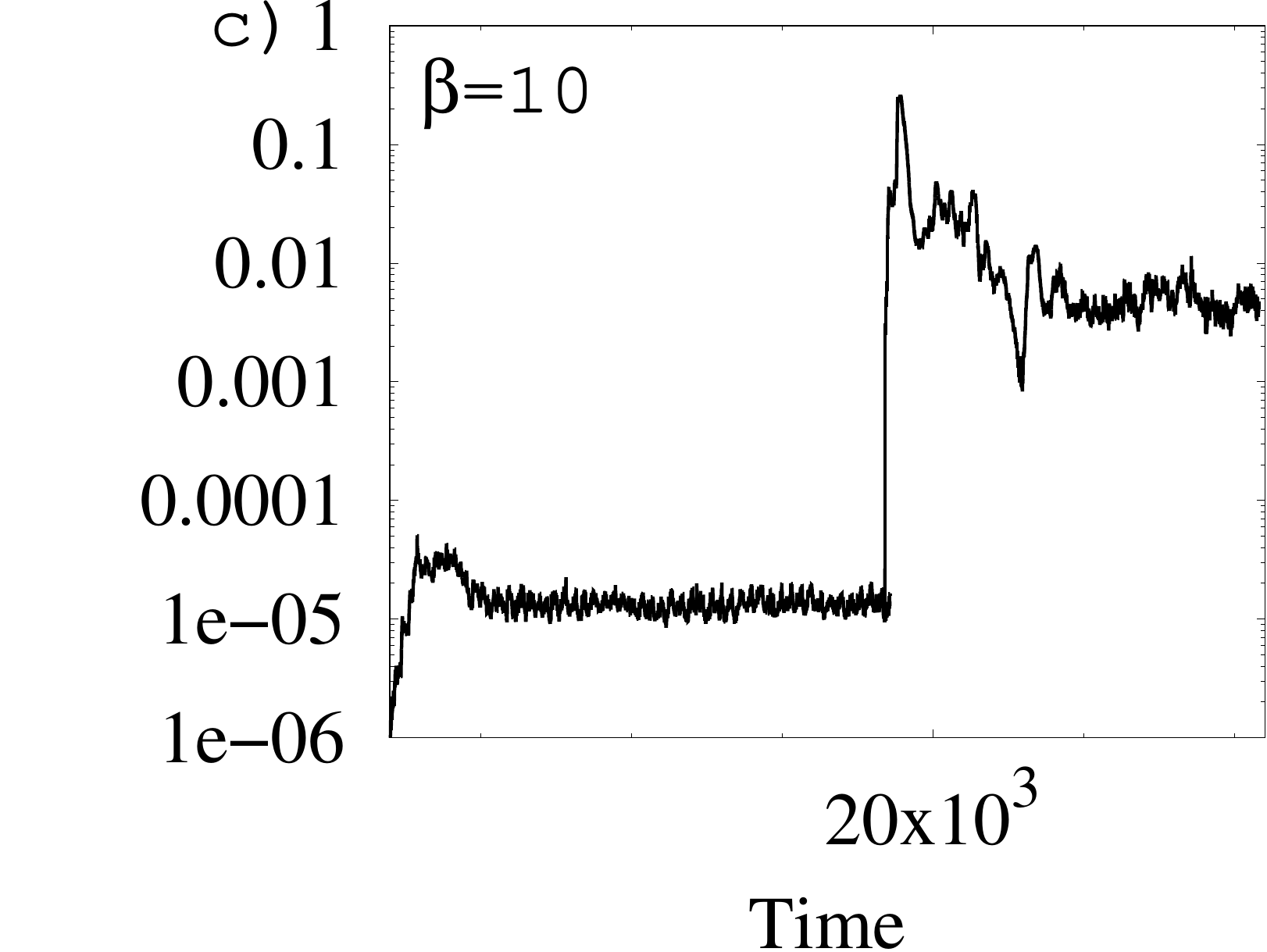}
\caption{Time variation of the ratio of the total mass outflow 
through the top and bottom z-boundaries to the mass inflow through the right radial boundary. 
Plasma betas are: (a) 50, (b) 25 and (c) 10 (cases A1, A2 and A3, respectively). The specific angular momentum of 1.65 
is used for all the cases shown here. We see that the mass outflow is significantly enhanced by the presence of 
magnetic field. Specifically, for plasma beta of 10 (strong field), nearly 25\% 
of injected matter can leave through the outer boundaries. The nature of the outflow 
is episodic and almost simultaneous with the time during which the flux tubes stay within the CENBOL.}
\label{fig7}
\end{figure*}

\begin{figure}
\includegraphics[width=82mm]{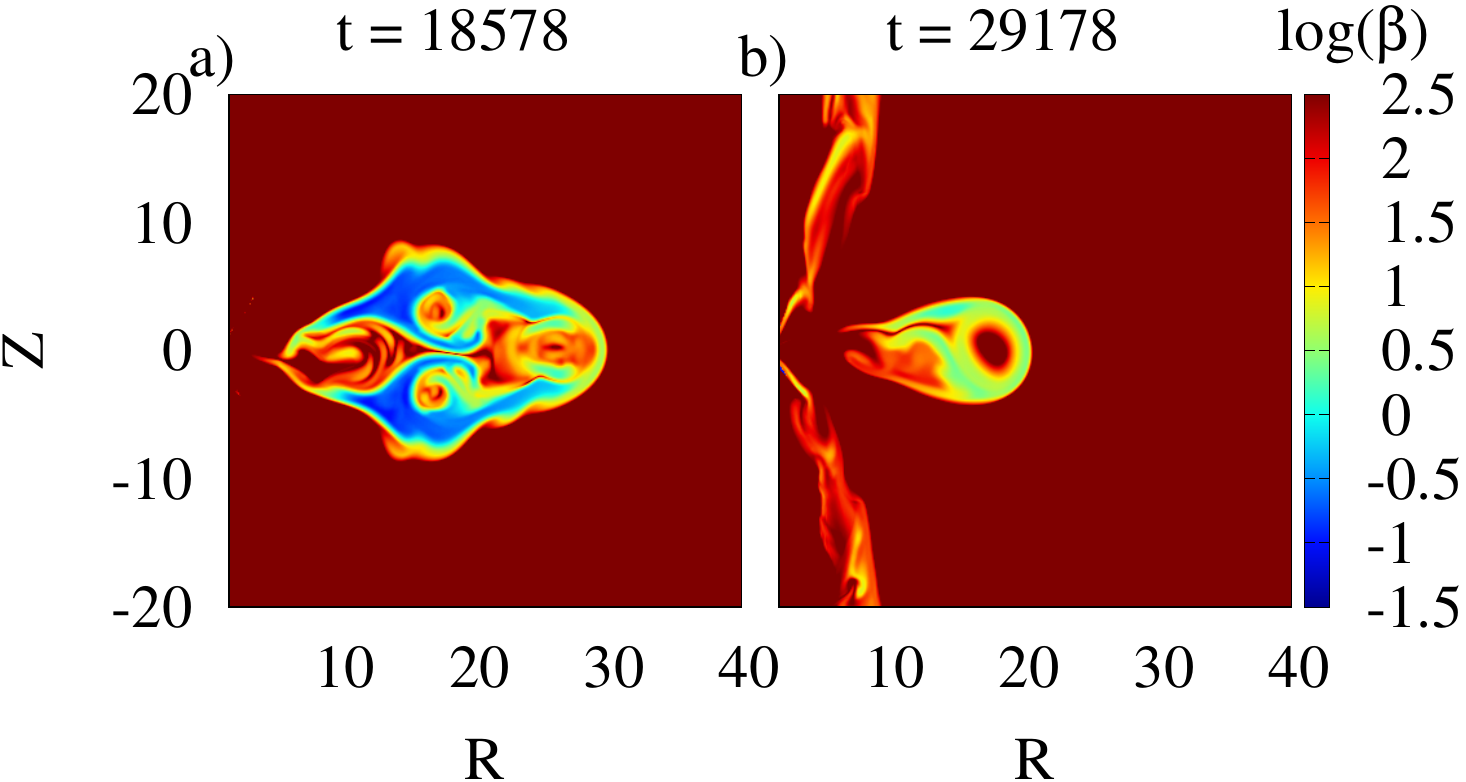}
\caption{Distribution of plasma beta ($\beta$) inside the accretion disk.
Figure has been drawn for run A2 and initial $\beta$ is 25 for this run.
a) shows $\beta$ after the field ropes arrived and compressed inside the CENBOL region.
Color bar shows that the $\beta$ value goes down below 0.1 which
clearly demonstrates that magnetic pressure increases by atleast 2 orders of
magnitude compared to the thermal pressure inside some part of CENBOL region.
b) shows $\beta$ at a later time when part of the magnetic flux left the CENBOl region
and the flow reached a steady state.}
\label{fig8}
\end{figure}

\begin{figure*}
\includegraphics[width=55mm]{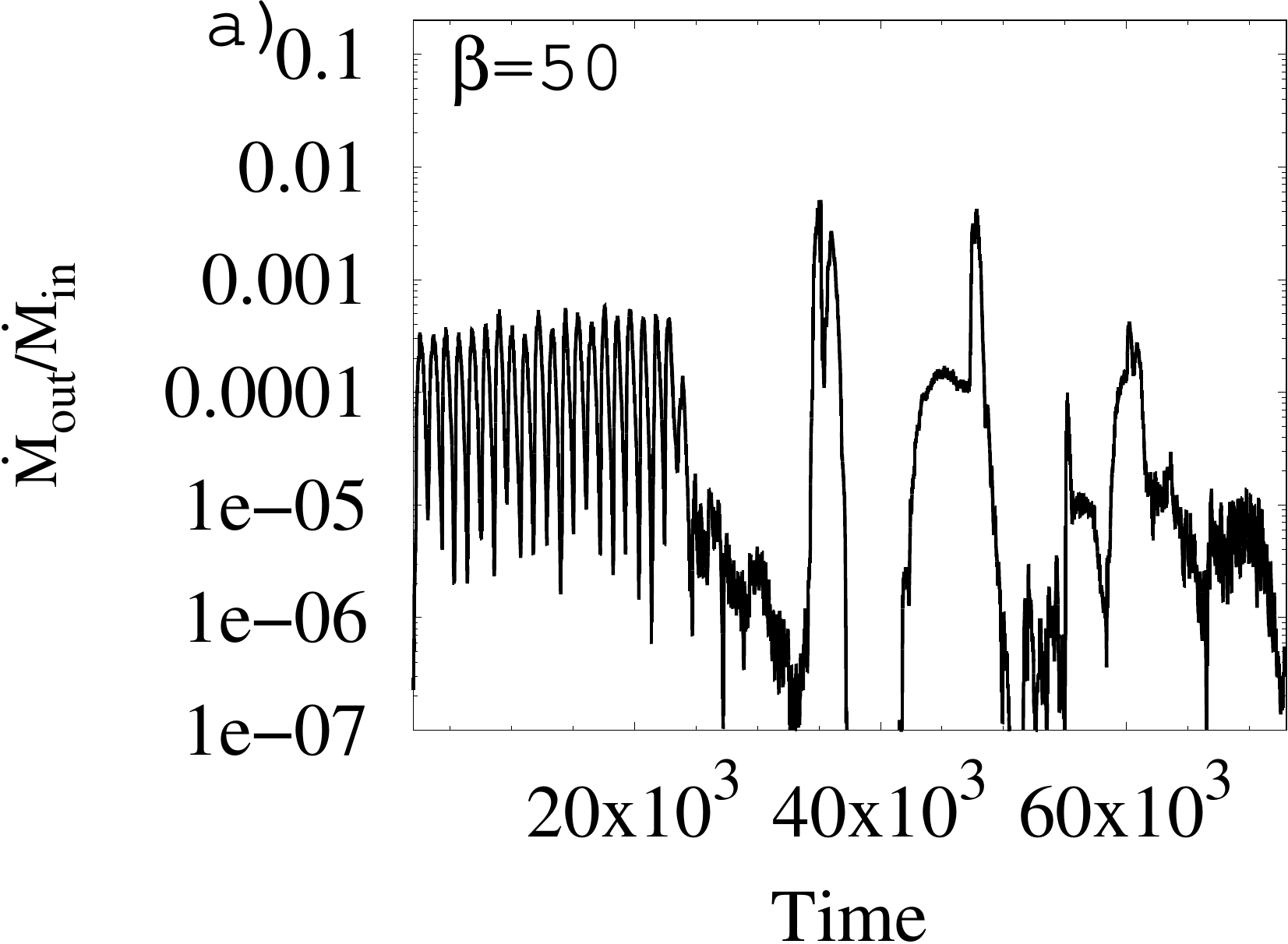}
\includegraphics[width=55mm]{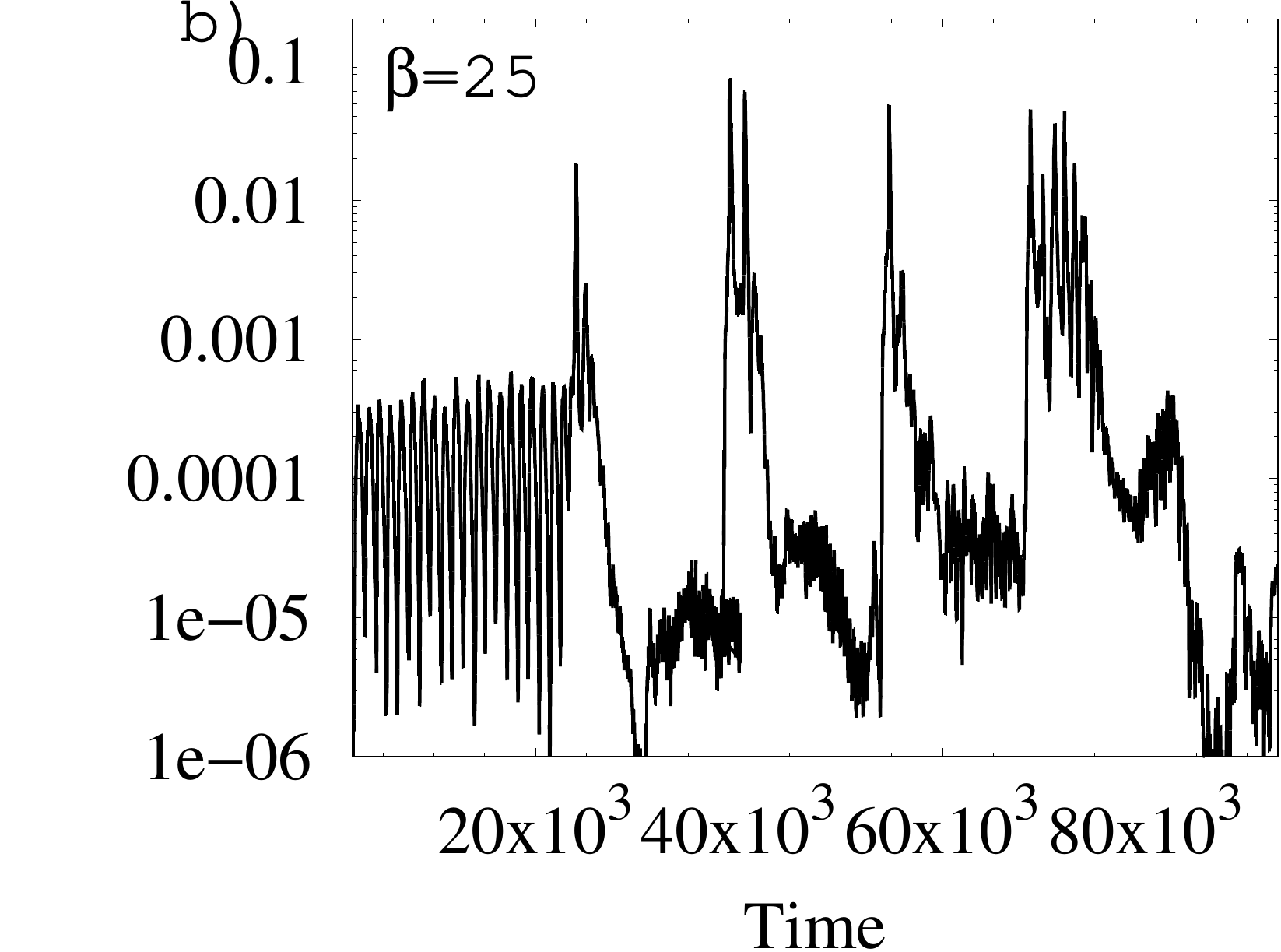}
\includegraphics[width=55mm]{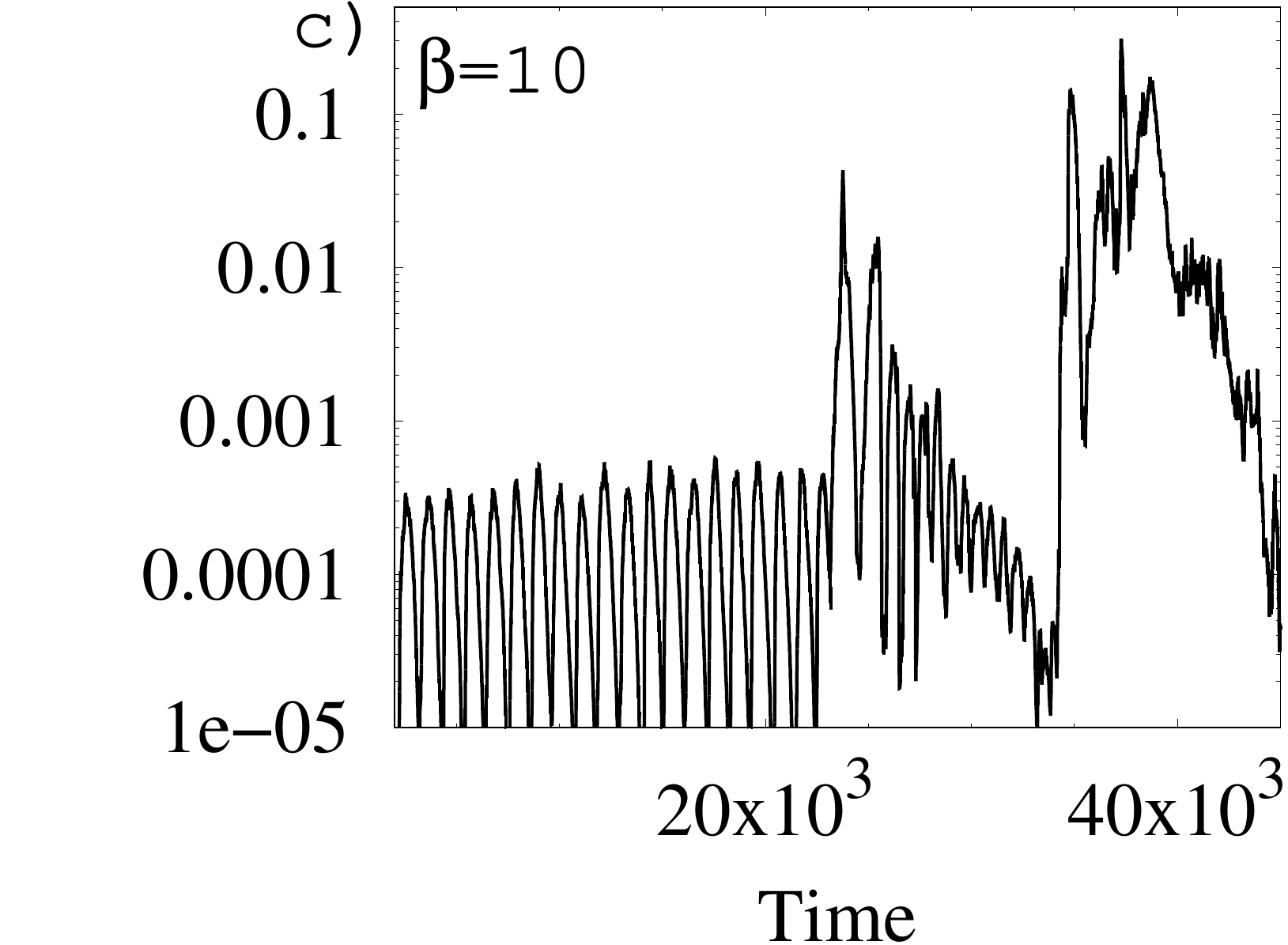}
\caption{\autoref{fig9} is same as \autoref{fig7} except that the specific angular momentum of $1.50$ is used (cases B1, B2 and B3 respectively). 
Here again, we can see that the mass outflow is significantly 
enhanced in presence of the magnetic field. Also, for the plasma beta of 10, nearly 25\% 
of injected matter can leave through the outer boundaries even for this lower angular 
momentum accretion flow. 
}
\label{fig9}
\end{figure*}

\autoref{fig7}(a-c) show time variation of the total outflow for cases A1, A2 and A3, respectively. 
On the y-axis, we plot the ratio of the total outflow rate ($\dot{M}_{\rm out}$) through 
the top and bottom Z-boundaries to the total inflow rate ($\dot{M}_{\rm in}$) through 
the right radial boundary. Since the specific angular momentum is low ($\lambda=1.65$),
we do not notice significant outflow during the initial un-magnetized accretion flow. 
This is because the outflow is centrifugally driven. In our simulations, 
the magnetic field passes through the standing strong magnetosonic shock. As a 
result, the post-shock field is significantly amplified, increasing its ability to drive outflows. 
Thus, as soon as the first flux rope arrive at the CENBOL region, 
the ratio of outflow to inflow is increased by 2-3 orders of magnitude. Specifically, 
for case A3 ($\beta=10$, i.e., strong magnetic field), nearly 25\% of the injected 
matter can leave through the outer boundaries. The enhancement of the magnetic pressure in the CENBOL due to
compression and as a result, it pushes the CENBOL boundary outward, increases 
the area of the base of the jet and drives a large fraction of inflow towards 
the vertical direction (see \autoref{fig8}). After the total pressure of the CENBOL region is reduced, 
the shock returns (\autoref{fig5}a-c) to a smaller radius and the outflow rate is also 
reduced before the next flux rope arrives. Thus the mass flux is also
controlled by the flux tube. It is to be noted that each flux rope, 
after being sheared by differential motion, stays inside the CENBOL for a 
substantial amount of time during which it drives the outflow. \autoref{fig9}(a-c) 
show the time variation of the outflow rate 
for cases B1, B2 and B3, respectively. Here, again, we notice a similar behavior 
as in the previous cases. The outflow rate is correlated with the shock location shown in \autoref{fig6}(a-c).

\begin{figure}
\includegraphics[width=\columnwidth]{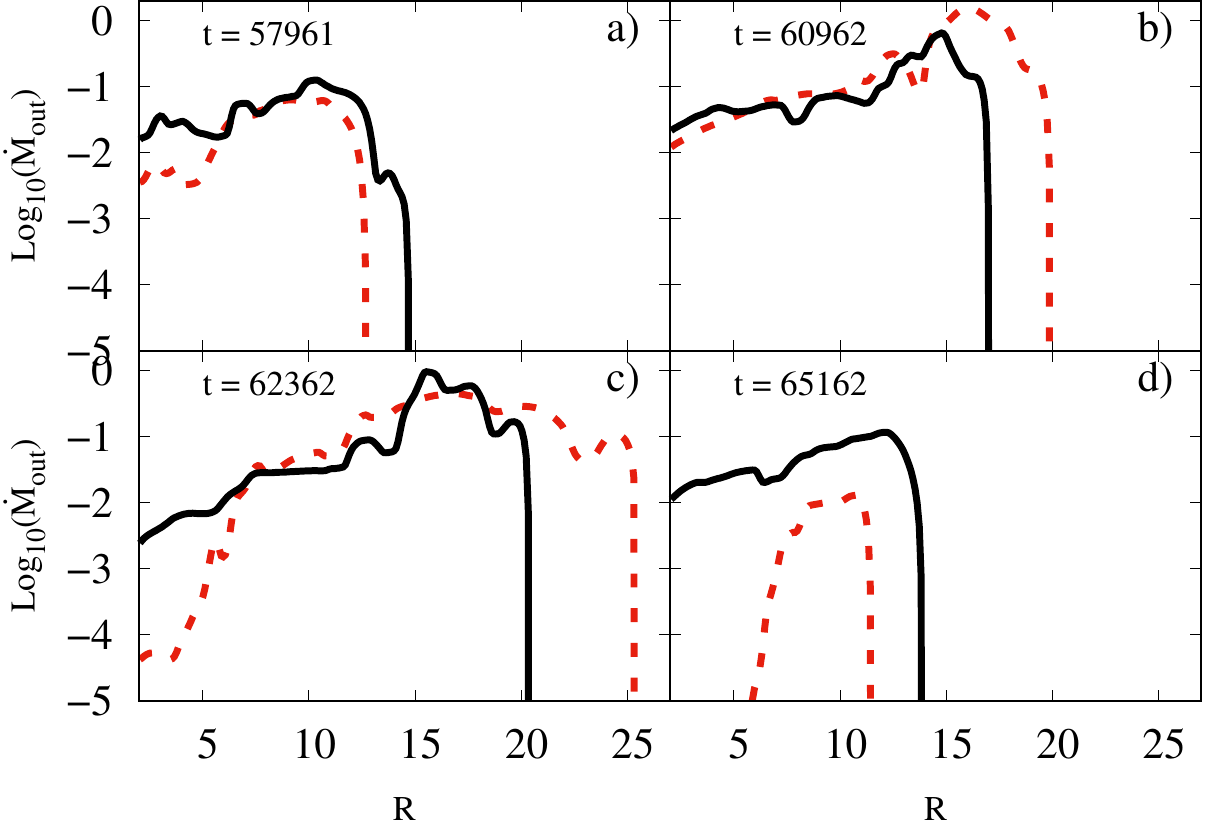}
\caption{Radial variation of the mass outflow through the upper (dashed red) 
and lower (solid black) z-boundaries at four different times marked by four arrows in \autoref{fig7}b. 
Mass outflow through both the z-boundaries, placed at 50 $r_g$ away 
from the black hole, takes place mostly through the region close to the axis. This shows 
the outflowing matter is not wind-like, rather, it forms a collimated jet. 
Specific angular momentum of 1.65 and a plasma beta of 25 is used (case A2). 
The jet power flip-flops in the upper and lower quadrants.
}
\label{fig10}
\end{figure}

We now focus on the collimation properties of the outflowing matter. We consider 
the case A2 for our analysis. \autoref{fig10} shows the radial variation of the mass outflow rate
($\dot{M}_{\rm out}$) through the upper (dashed red) and lower (solid black) Z-boundaries 
located at z = 50 and z = -50, respectively,
at four different times marked by four red arrows in \autoref{fig7}b. \autoref{fig10}a shows the outflow 
rate just before the initiation of the fourth flux rope injection. 
By this time, the third  flux rope has passed the CENBOL region 
and the disc has achieved a stable configuration. The outflow through both the 
Z-boundaries is found be taking place close to the axis of rotation. 
\autoref{fig10}(b-c) depict the outflow just after the flux rope has entered the CENBOL. 
The outflow rate is found to be increased by an order of magnitude, which is expected. 
Even for these two configurations, the outflow is found to be taking place mostly 
close to the axis of rotation. Finally, \autoref{fig10}d shows the stabilized state and again, 
the rate is higher close to the axis. All these Figures show that outflowing 
matter is not wind-like with high opening angle, rather, the matter leaves the disc in a jet-like collimated fashion.

\begin{figure}
\includegraphics[width=\columnwidth]{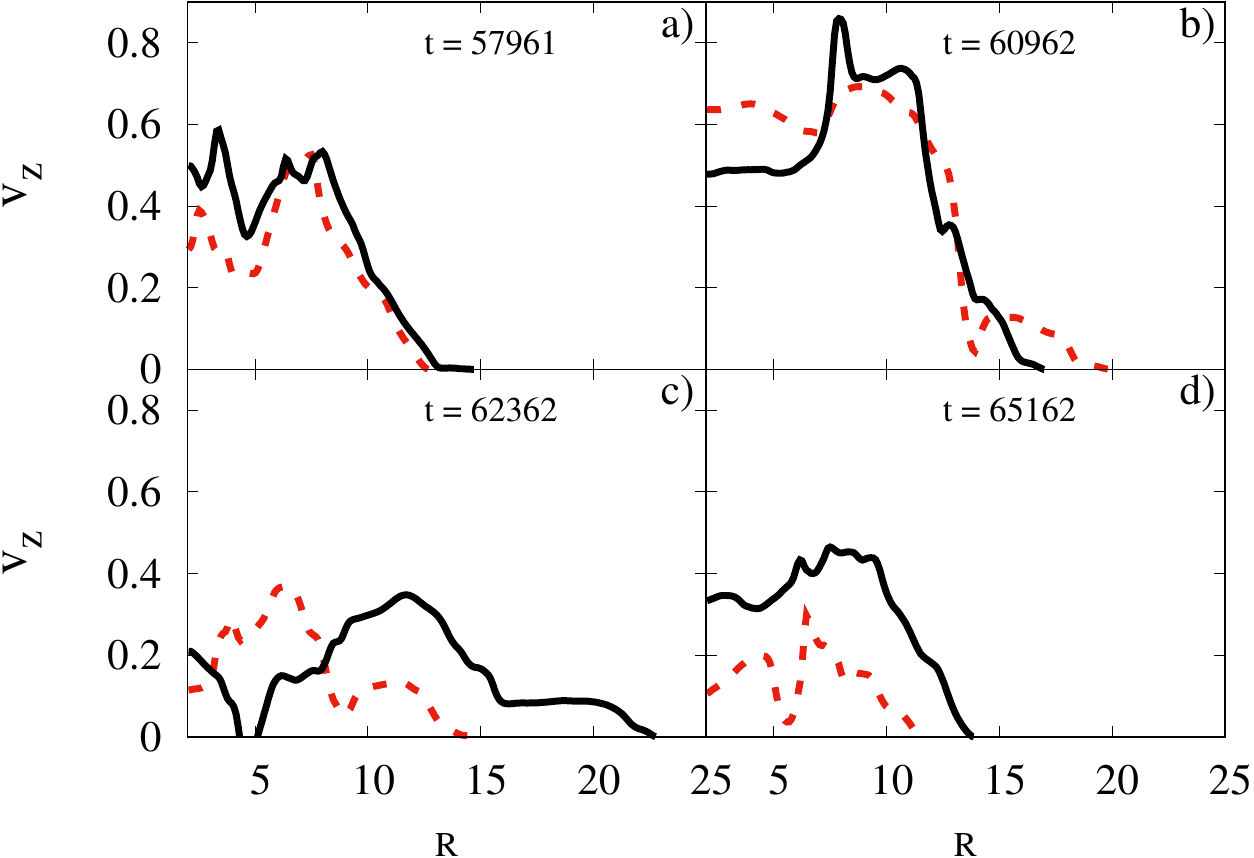}
\caption{Radial variation of the z-component of velocity of the 
outflowing matter through the upper (dashed red) and lower (solid black) z-boundaries 
at four different times same as in \autoref{fig10}. For lower boundary, the sign of $V_z$ has 
been flipped for the purpose of comparison. We see that the outflowing matter 
has the higher ejection velocity close to the axis, showing the collimated nature of the outflow.
}
\label{fig11}
\end{figure}

To further prove the collimation of the outflow, in \autoref{fig11}, we plot the radial variation 
of $V_z$ of the matter flowing through the upper (dashed red) and lower (solid black) 
Z-boundaries at the same four times as in \autoref{fig10}. Outflowing matter through the upper 
Z-boundary has a positive velocity, whereas the same through the lower Z-boundary 
has a negative velocity. The sign of $V_z$ has been flipped while plotting 
for the lower Z-boundary. These plots show that the higher velocity outflowing matter 
truly leaves the computational domain very close to the axis, showing the collimated nature of the outflow.

\begin{figure*}
\begin{center}
\includegraphics[width=120mm]{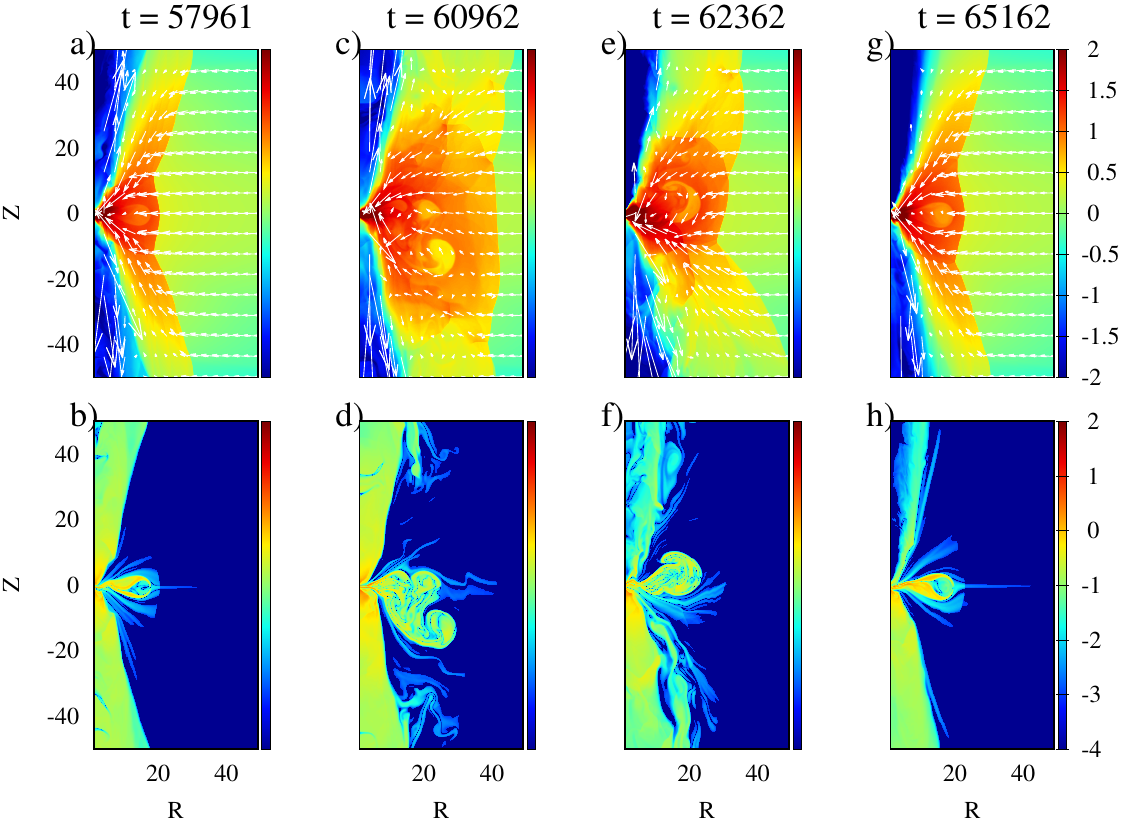}
\end{center}
\caption{Overall picture of the simulation flow close to the black hole where 
we see the inflow, outflow and the $B_\phi$ distribution 
inside the disc. (a), (c), (e) and (g) show the density distribution on log scale, 
over-plotted with velocity vectors, at four different times as in \autoref{fig10} and \autoref{fig11}. 
Figures (b), (d), (f) and (h) show the absolute value of $B_\phi$, in log scale, 
at the same times. We see that the outflowing matter with very high velocities 
are bound by the toroidal component of the magnetic field. In other words, $B_\phi$ 
helps to collimate the outflow. The specific angular momentum of 1.65 and a 
plasma beta of 25 is used for this case (run A2).
}
\label{fig12}
\end{figure*}

In order to explore the reason for the collimation, we investigate the correlation 
between the outflow and the toroidal component of magnetic field ($B_\phi$).
\autoref{fig12}a, c, e and g show the density distribution on log scale, over-plotted 
with velocity vectors, at four different times as in \autoref{fig7} and \autoref{fig9}. 
\autoref{fig12}b, d, f and h show the absolute value of $B_\phi$, again on log scale, 
at the same times. We clearly see that the outflowing matter with very 
high velocities are bound by the toroidal component of the magnetic field. 
Thus, $B_\phi$ helps to collimate the outflow due to the so-called `hoop-stress' (see also, 
CD94 and DC94). 

\begin{figure*}
\begin{center}
\includegraphics[width=140mm]{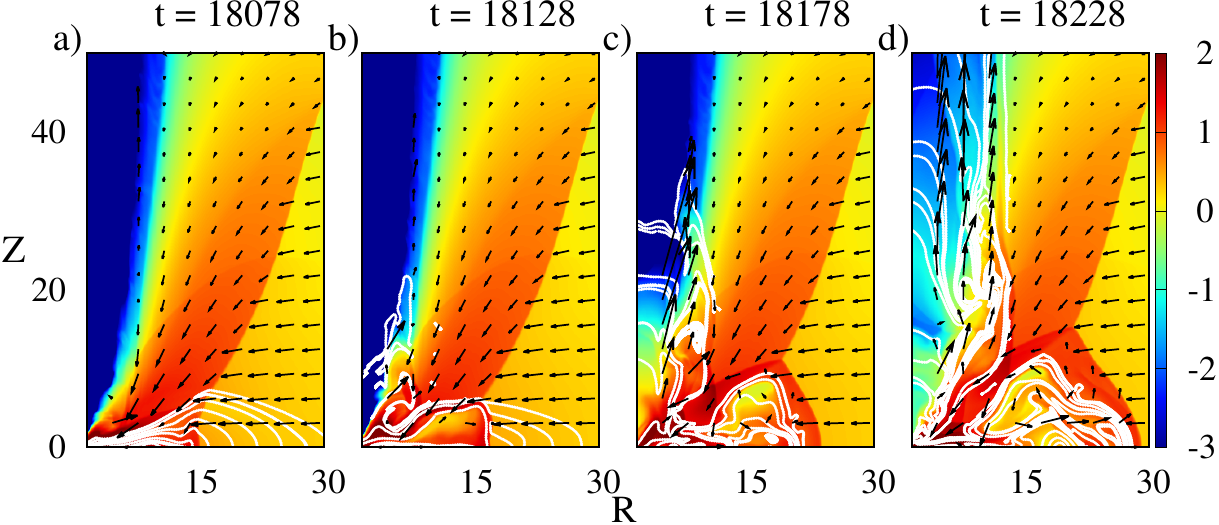}
\end{center}
\caption{Escape of magnetic flux from the disk is shown. Zoomed in snapshots at four
different times (marked on the top of each snapshot) in the upper quadrant are shown here.
The color shows the logarithm of density. White lines show the poloidal magnetic field lines and
black arrows show the velocity vectors. This Figure has been drawn for run A3.
}
\label{fig13}
\end{figure*}

\autoref{fig13} shows how magnetic flux escapes from the disk. We have shown the zoomed in
snapshots at four different times (marked on the top of each snapshot) in the upper quadrant
of the computational domain for runA3. Poloidal magnetic field lines are shown by white lines
and the velocity vectors of the matter are shown by the black arrows. The magnetic flux is seen
to be escaping through the initially empty funnel area. Incoming poloidal field loops
first enter the low density area close to the axis and subsequently expand to form 
large scale poloidal field lines. Length of the velocity vectors confirms that the matter
in this region also achieves high velocity parallel to the magnetic field lines.

\subsection{On the question of anchoring of the flux tubes}

In stars with radiative core and convective envelope it is well known that the magnetic 
fields are  anchored in between these two strata and may time to time be buoyant and 
partly leave the convective envelope creating structured corona. Random motions of the 
convective zones often reconnect the flux tubes and the magnetic energy heats up the corona. 
In the case of accretion discs, it is customary to assume a corona without proving that 
magnetic field lines could be anchored. Preliminary study of this by CD94 and DC94 
suggest that only if the entropy gradient is favorable, as in the stars, sometimes 
the flux tubes will be trapped and oscillate. In the present context we have seen 
the oscillation of the flux tubes inside the CENBOL as well. Thus instead of leaving 
the disc as a whole, the torn flux tube leave part by part, in the form of toroidal loops. 
As in stars, the possibility of this behaviour depends on the entropy gradient. 
In stars, for stability reason, entropy must go down towards the surface as well as towards the center. 

\begin{figure*}
\includegraphics[width=160mm]{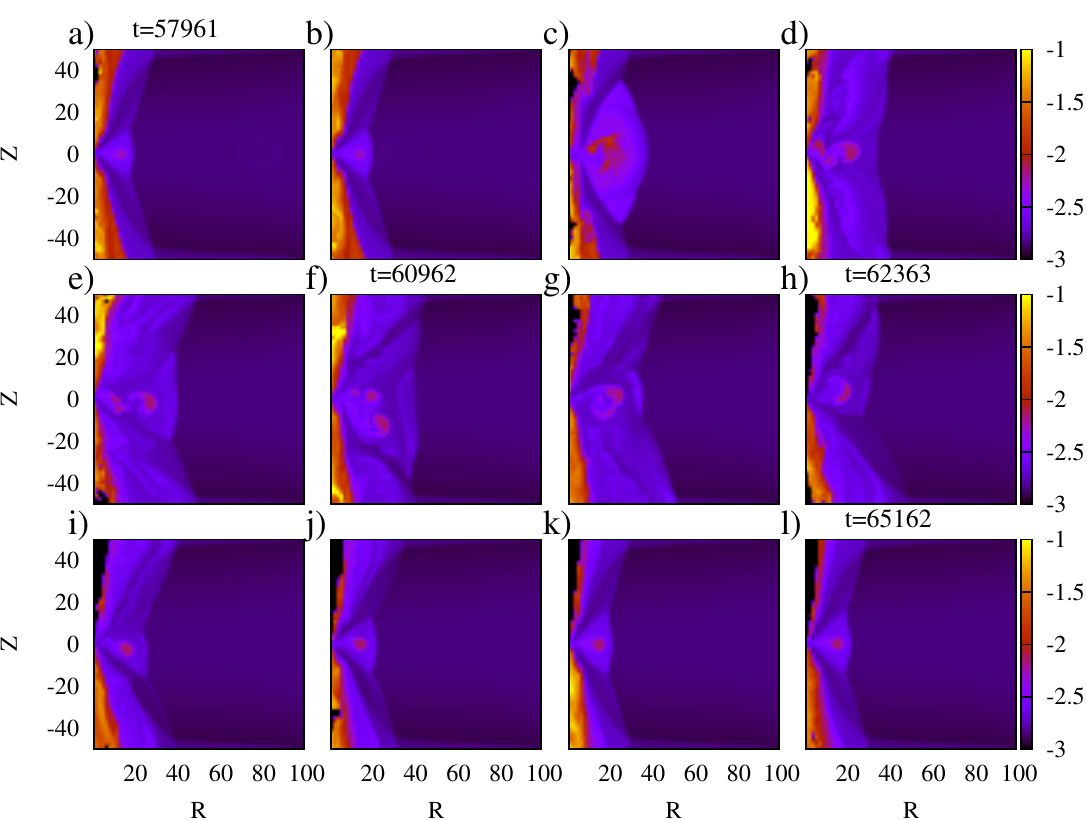}
\caption{Entropy ($P/\rho^\gamma$) distribution  in log scale at 
nearly equidistant times between t=57961 and 65162. The panels marked "t=xxxx" 
are drawn at the same time as in \autoref{fig10}-\autoref{fig12}. The extra Figures are added to 
get better understanding of the time evolution of entropy. 
The zones are re-binned on a 16x16 mesh in order to have a coarse grid data.
In the black regions, the mass density is less than 1\% of the injected density.}

\label{fig14}
\end{figure*}

\begin{figure*}
\includegraphics[width=160mm]{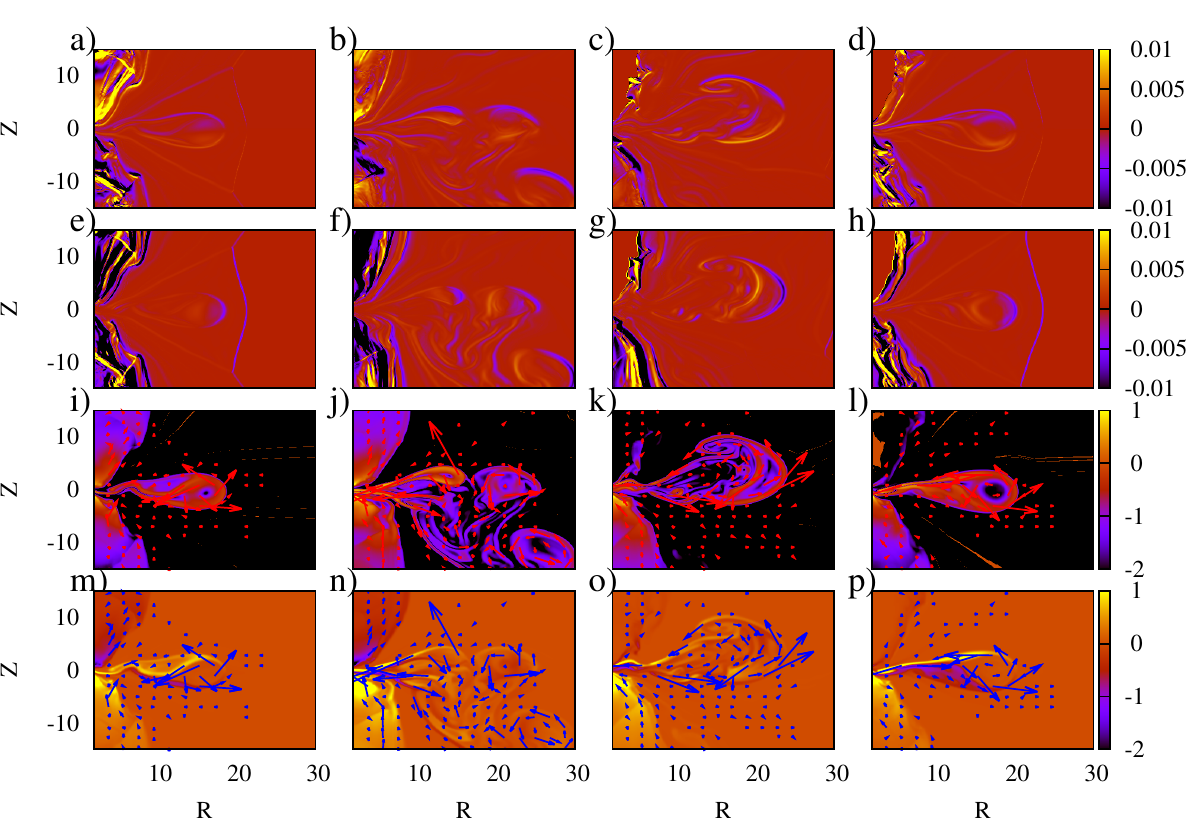}
\caption{First row  (a, b, c, d) shows the dS/dZ distribution at four different times 
(t=57961, 60962, 62363 and 65162). Second row (e, f, g, h) shows dS/dR at the same times. 
The color scales are chosen so that the variation is visible. 
Third row (i,j,k,l) shows $\sqrt(B_R^2 + B_\phi^2 + B_Z^2)$ distribution in log scale at those times.
Fourth row (m,n,o,p) shows the distribution of $B_\phi$ in linear scale. Here the lower limit 
of the color scale has been set to -2, but the upper limit represents true maximum value of the data. 
Vectors represent ($B_R,B_Z$) field and their lengths are proportional to $\sqrt(B_R^2 + B_Z^2)$.
}
\label{fig15}
\end{figure*}

In \autoref{fig14}(a-l), we present the entropy map of the flow at roughly equidistant time during 
the interval of our interest: $t=57961$ to $65162$. 16x16 course grid is chosen to avoid 
very small scale variations. The color scale shows that a high entropy blob expands 
in all direction and then collapses. This is the behaviour of the shock location as well. 
Second: highest entropy flow forms along the axis where the matter escapes after passing 
through the CENBOL. 
In \autoref{fig15}(a-p), we present the vertical and radial entropy gradients at the same times 
when \autoref{fig10}-\autoref{fig12} were drawn in a zoomed region close to the black hole. The rows from 
top to bottom show: in colors respectively the (a-d): entropy gradients in the vertical 
direction (dS/dZ), (e-h): radial gradient (dS/dR), (i-l): the magnitude of the net field 
$\sqrt(B_R^2 + B_\phi^2 + B_Z^2)$ in log scale and (m-p): the magnitude of the $B_\phi$ 
component in log scale. The corresponding color scales are placed beside. The arrows 
represent the field direction in $R,Z$ plane. The lengths are proportional to $\sqrt(B_R^2 + B_Z^2)$. 
In the black regions, the mass density is less than 1\% of the injected density.
We note very interesting behaviour: the vertical gradient is 
very strongly negative in the upper hemisphere and weakly positive in the lower hemisphere. 
This drives the magnetic blob upwards. Similarly, the radial gradient is negative at the 
forefront of the expanding bubble. This also drives the flux tube away from the axis, 
except that due to resistance of the inflow upstream the field  mostly escapes vertically. 
From the arrows, we see a strong helicity in the flux rope. We also note that $B_\phi$ 
is strong in the lower jet causing a strong expulsion of fluxes along the axis. 
So, while inside the disc, the flux is driving towards the upper hemisphere, inside 
the jet the field (from previous episodes) already escapes through the lower jet. 
We generally find the funnel region along the axis is magnetically very active and 
high degree of reconnection cannot be ruled out. When such energy dissipation is 
included in the code in future, we anticipate that a higher outflow rate would be seen.

\section{Conclusions}
\label{sec:sec4}

In this paper, we have studied several aspects of a rotating transonic (advective) flow in presence of 
magnetic flux tubes. In the literature, much of the studies have been made only with hydrodynamic flows.
These studies showed, both theoretically as well as with numerical simulations, how standing shocks 
could form in between two sonic points during its passages towards the black hole. However, in a realistic
accretion, significant magnetic activity could be present which might disrupt the flow or convert the flow to 
a Keplerian disc in case the magnetic transport of angular momentum were large enough. Also, the magnetic fields 
may also contribute to accelerate and collimate matter in the outflow. Thus it is very important to study the
effects of magnetic fields on the accretion and outflows. Following CD9 and DC94, \citet{deb2017}
recently showed that a single flux tube may be able to collimate the jet and outflow coming
out of the post-shock region. However they used a hydrodynamic code which treated the fields passively. 

In the present paper, we use an ideal MHD code and applied it around a Schwarzschild black hole 
geometry. As a starting point, we wanted to study the effects of toroidal flux tubes which were injected 
from the outer boundary. We assumed the random magnetic fields to be sheared predominantly due to the rotational
velocity close to the black hole and the fields are predominantly toroidal in shape.
These flux tubes could have various strengths or cross-sections. However, we select only a single 
cross-section. Similarly, we also injected these field lines one by one, while in reality, the situation 
could be more complex and many field lines may be entering simultaneously. 

Even with this simple consideration, we established a few very important results. First of all, we show
that the flux tube strengthen the formation of the shockwave and repeated bombardments of flux tube do not
destabilize the shock front or the flow geometry at all. This is important, since the post-shock region behaves as the 
hot Compton cloud which inverse Comptonizes the soft photons and produce a power-law photon in a typical
disc spectrum \citep{chakraba1997,gcl2009,ggcl2010,ggc2012,ggc2014}.
Second, the flux tubes are sheared as they propagate close to the 
black hole and are ejected along vertical axis. The funnel region is found to have high entropy and is also magnetically very 
active. The outflows are found to be collimated with high velocity components. We also notice that 
along the direction in which entropy decreases, the flux tubes escape, very similar
to what happens in a star with a radiative core and convective envelope (like our Sun). However, they 
were not seen to form corona as on the solar surface since the regions of highest entropy are highly dynamic
and anchoring flux tubes become difficult. 

In presence of magnetic flux tubes, the field pressure adds up with the thermal pressure inside the CENBOL 
region making the shock location at  a higher radius. The shock ejects the flux tubes oscillating back and forth
thereby modulating the outflows as well. Furthermore, we find that the flux tube is ejected 
asymmetrically in the upper and the lower quadrants. 

A phenomenon discovered by us may be important:  we find that typically the matter losses angular momentum at the 
inner-half inside the flux tube and gains it at the outer half. They are radially stretched by this 
process and also due to differential radial velocity. Thus the flux tubes may be used to transport
angular momentum inside the disc, as well.

In the literature, use has been made of large scale magnetic field lines for jet formation and collimation. While 
it is unclear, how these field lines are generated, we find that generically obtained flux tubes such as those 
we use here can perform equally well in acceleration and collimation. Rapid expulsion of flux tubes 
along the vertical axis may also eject relativistic blobs of matter. Thus both the compact and blobby jets are 
possible in our configuration. These aspects and the effects of such magnetized flows on disc 
spectrum will be studied in future and will be reported elsewhere.

\acknowledgments

DSB acknowledges support via NSF grants NSF-ACI-1533850, NSF-DMS-1622457, NSF-ACI-1713765.
Several simulations were performed on a cluster at UND that is run by the Center for Research Computing.
Computer support on NSF's XSEDE and Blue Waters computing resources is also acknowledged.
SKC acknowledges a grant from Higher Education Dept. of the State of West Bengal.
JK acknowledges the support by Basic Science Research Program through the National
Research Foundation of Korea (NRF) funded by the Ministry of Education (2018R1D1A1B07042949) and National
Supercomputing Center with supercomputing resources including technical support (KSC-2018-CRE-0098).

%

\software{RIEMANN code
\citep{bal1998a,bal1998b,bal2004,bal2009,bal_spi1999a,bal_spi1999b,brdm2009,bal2013}}


\bibliographystyle{aasjournal.bst}
\bibliography{ref} 

\begin{thebibliography}{}
\expandafter\ifx\csname natexlab\endcsname\relax\def\natexlab#1{#1}\fi
\providecommand{\url}[1]{\href{#1}{#1}}
\providecommand{\dodoi}[1]{doi:~\href{http://doi.org/#1}{\nolinkurl{#1}}}
\providecommand{\doeprint}[1]{\href{http://ascl.net/#1}{\nolinkurl{http://ascl.net/#1}}}
\providecommand{\doarXiv}[1]{\href{https://arxiv.org/abs/#1}{\nolinkurl{https://arxiv.org/abs/#1}}}

\bibitem[{{Balbus} \& {Hawley}(1991)}]{balbus1991}
{Balbus}, S.~A., \& {Hawley}, J.~F. 1991, \apj, 376, 214,
  \dodoi{10.1086/170270}

\bibitem[{{Balsara}(1998{\natexlab{a}})}]{bal1998a}
{Balsara}, D.~S. 1998{\natexlab{a}}, \apjs, 116, 133, \dodoi{10.1086/313093}

\bibitem[{{Balsara}(1998{\natexlab{b}})}]{bal1998b}
---. 1998{\natexlab{b}}, \apjs, 116, 119, \dodoi{10.1086/313092}

\bibitem[{{Balsara}(2004)}]{bal2004}
---. 2004, \apjs, 151, 149, \dodoi{10.1086/381377}

\bibitem[{{Balsara}(2009)}]{bal2009}
---. 2009, Journal of Computational Physics, 228, 5040,
  \dodoi{10.1016/j.jcp.2009.03.038}

\bibitem[{{Balsara} {et~al.}(2013){Balsara}, {Meyer}, {Dumbser}, {Du}, \&
  {Xu}}]{bal2013}
{Balsara}, D.~S., {Meyer}, C., {Dumbser}, M., {Du}, H., \& {Xu}, Z. 2013,
  Journal of Computational Physics, 235, 934, \dodoi{10.1016/j.jcp.2012.04.051}

\bibitem[{{Balsara} {et~al.}(2009){Balsara}, {Rumpf}, {Dumbser}, \&
  {Munz}}]{brdm2009}
{Balsara}, D.~S., {Rumpf}, T., {Dumbser}, M., \& {Munz}, C.-D. 2009, Journal of
  Computational Physics, 228, 2480, \dodoi{10.1016/j.jcp.2008.12.003}

\bibitem[{{Balsara} \& {Spicer}(1999{\natexlab{a}})}]{bal_spi1999b}
{Balsara}, D.~S., \& {Spicer}, D. 1999{\natexlab{a}}, Journal of Computational
  Physics, 148, 133, \dodoi{10.1006/jcph.1998.6108}

\bibitem[{{Balsara} \& {Spicer}(1999{\natexlab{b}})}]{bal_spi1999a}
{Balsara}, D.~S., \& {Spicer}, D.~S. 1999{\natexlab{b}}, Journal of
  Computational Physics, 149, 270, \dodoi{10.1006/jcph.1998.6153}

\bibitem[{{Blandford} \& {Payne}(1982)}]{bp1982}
{Blandford}, R.~D., \& {Payne}, D.~G. 1982, \mnras, 199, 883,
  \dodoi{10.1093/mnras/199.4.883}

\bibitem[{{Camenzind}(1989)}]{camen1989}
{Camenzind}, M. 1989, in Astrophysics and Space Science Library, Vol. 156,
  Accretion Disks and Magnetic Fields in Astrophysics, ed. G.~{Belvedere}, 129,
  \dodoi{10.1007/978-94-009-2401-7_14}

\bibitem[{{Chakrabarti}(1986)}]{chakraba1986}
{Chakrabarti}, S.~K. 1986, \apj, 303, 582, \dodoi{10.1086/164104}

\bibitem[{{Chakrabarti}(1989)}]{chakraba1989}
---. 1989, \apjl, 337, L89, \dodoi{10.1086/185385}

\bibitem[{{Chakrabarti}(1990)}]{chakraba1990}
---. 1990, {Theory of Transonic Astrophysical Flows}, \dodoi{10.1142/1091}

\bibitem[{{Chakrabarti}(1997)}]{chakraba1997}
---. 1997, \apj, 484, 313, \dodoi{10.1086/304325}

\bibitem[{{Chakrabarti} {et~al.}(2004){Chakrabarti}, {Acharyya}, \&
  {Molteni}}]{cam2004}
{Chakrabarti}, S.~K., {Acharyya}, K., \& {Molteni}, D. 2004, \aap, 421, 1,
  \dodoi{10.1051/0004-6361:20034523}

\bibitem[{{Chakrabarti} \& {Bhaskaran}(1992)}]{chakraba1992}
{Chakrabarti}, S.~K., \& {Bhaskaran}, P. 1992, \mnras, 255, 255,
  \dodoi{10.1093/mnras/255.2.255}

\bibitem[{{Chakrabarti} \& {D'Silva}(1994)}]{cd1994}
{Chakrabarti}, S.~K., \& {D'Silva}, S. 1994, \apj, 424, 138,
  \dodoi{10.1086/173878}

\bibitem[{{Coroniti}(1981)}]{coro1981}
{Coroniti}, F.~V. 1981, \apj, 244, 587, \dodoi{10.1086/158739}

\bibitem[{{Deb} {et~al.}(2017){Deb}, {Giri}, \& {Chakrabarti}}]{deb2017}
{Deb}, A., {Giri}, K., \& {Chakrabarti}, S.~K. 2017, \mnras, 472, 1259,
  \dodoi{10.1093/mnras/stx1721}

\bibitem[{{D'Silva} \& {Chakrabarti}(1994)}]{dsilva1994}
{D'Silva}, S., \& {Chakrabarti}, S.~K. 1994, \apj, 424, 149,
  \dodoi{10.1086/173879}

\bibitem[{{Eardley} \& {Lightman}(1975)}]{el1975}
{Eardley}, D.~M., \& {Lightman}, A.~P. 1975, \apj, 200, 187,
  \dodoi{10.1086/153777}

\bibitem[{{Eggum} {et~al.}(1985){Eggum}, {Coroniti}, \& {Katz}}]{eck1985}
{Eggum}, G.~E., {Coroniti}, F.~V., \& {Katz}, J.~I. 1985, \apjl, 298, L41,
  \dodoi{10.1086/184563}

\bibitem[{{Fukue}(1982)}]{fukue1982}
{Fukue}, J. 1982, \pasj, 34, 163

\bibitem[{{Galeev} {et~al.}(1979){Galeev}, {Rosner}, \& {Vaiana}}]{galeev1979}
{Galeev}, A.~A., {Rosner}, R., \& {Vaiana}, G.~S. 1979, \apj, 229, 318,
  \dodoi{10.1086/156957}

\bibitem[{{Garain} {et~al.}(2012){Garain}, {Ghosh}, \& {Chakrabarti}}]{ggc2012}
{Garain}, S.~K., {Ghosh}, H., \& {Chakrabarti}, S.~K. 2012, \apj, 758, 114,
  \dodoi{10.1088/0004-637X/758/2/114}

\bibitem[{{Garain} {et~al.}(2014){Garain}, {Ghosh}, \& {Chakrabarti}}]{ggc2014}
---. 2014, \mnras, 437, 1329, \dodoi{10.1093/mnras/stt1969}

\bibitem[{{Ghosh} {et~al.}(2009){Ghosh}, {Chakrabarti}, \& {Laurent}}]{gcl2009}
{Ghosh}, H., {Chakrabarti}, S.~K., \& {Laurent}, P. 2009, International Journal
  of Modern Physics D, 18, 1693, \dodoi{10.1142/S0218271809015242}

\bibitem[{{Ghosh} {et~al.}(2010){Ghosh}, {Garain}, {Chakrabarti}, \&
  {Laurent}}]{ggcl2010}
{Ghosh}, H., {Garain}, S.~K., {Chakrabarti}, S.~K., \& {Laurent}, P. 2010,
  International Journal of Modern Physics D, 19, 607,
  \dodoi{10.1142/S0218271810016555}

\bibitem[{{Giri} \& {Chakrabarti}(2013)}]{giri2013}
{Giri}, K., \& {Chakrabarti}, S.~K. 2013, \mnras, 430, 2836,
  \dodoi{10.1093/mnras/stt087}

\bibitem[{{Giri} {et~al.}(2010){Giri}, {Chakrabarti}, {Samanta}, \&
  {Ryu}}]{giri2010}
{Giri}, K., {Chakrabarti}, S.~K., {Samanta}, M.~M., \& {Ryu}, D. 2010, \mnras,
  403, 516, \dodoi{10.1111/j.1365-2966.2009.16147.x}

\bibitem[{{Hawley}(2000)}]{hawley2000}
{Hawley}, J.~F. 2000, \apj, 528, 462, \dodoi{10.1086/308180}

\bibitem[{{Hawley} \& {Balbus}(1991)}]{hawley1991}
{Hawley}, J.~F., \& {Balbus}, S.~A. 1991, \apj, 376, 223,
  \dodoi{10.1086/170271}

\bibitem[{{Hawley} \& {Krolik}(2001)}]{hk2001}
{Hawley}, J.~F., \& {Krolik}, J.~H. 2001, \apj, 548, 348,
  \dodoi{10.1086/318678}

\bibitem[{{Heyvaerts} \& {Norman}(1989)}]{heyv1989}
{Heyvaerts}, J., \& {Norman}, C. 1989, \apj, 347, 1055, \dodoi{10.1086/168195}

\bibitem[{{Koide} {et~al.}(1999){Koide}, {Shibata}, \& {Kudoh}}]{koide1999}
{Koide}, S., {Shibata}, K., \& {Kudoh}, T. 1999, \apj, 522, 727,
  \dodoi{10.1086/307667}

\bibitem[{{Konigl}(1989)}]{konigl1989}
{Konigl}, A. 1989, \apj, 342, 208, \dodoi{10.1086/167585}

\bibitem[{{Kudoh} {et~al.}(2002){Kudoh}, {Matsumoto}, \& {Shibata}}]{kudoh2002}
{Kudoh}, T., {Matsumoto}, R., \& {Shibata}, K. 2002, \pasj, 54, 267,
  \dodoi{10.1093/pasj/54.2.267}

\bibitem[{{Lee} {et~al.}(2016){Lee}, {Chattopadhyay}, {Kumar}, {Hyung}, \&
  {Ryu}}]{lee2016}
{Lee}, S.-J., {Chattopadhyay}, I., {Kumar}, R., {Hyung}, S., \& {Ryu}, D. 2016,
  \apj, 831, 33, \dodoi{10.3847/0004-637X/831/1/33}

\bibitem[{{Lovelace}(1976)}]{lovelace1976}
{Lovelace}, R.~V.~E. 1976, \nat, 262, 649, \dodoi{10.1038/262649a0}

\bibitem[{{Lynden-Bell}(1978)}]{lynden1978}
{Lynden-Bell}, D. 1978, \physscr, 17, 185, \dodoi{10.1088/0031-8949/17/3/009}

\bibitem[{{McKinney}(2006)}]{mckinney2006}
{McKinney}, J.~C. 2006, \mnras, 368, 1561,
  \dodoi{10.1111/j.1365-2966.2006.10256.x}

\bibitem[{{Molteni} {et~al.}(1994){Molteni}, {Lanzafame}, \&
  {Chakrabarti}}]{mlc1994}
{Molteni}, D., {Lanzafame}, G., \& {Chakrabarti}, S.~K. 1994, \apj, 425, 161,
  \dodoi{10.1086/173972}

\bibitem[{{Molteni} {et~al.}(1996{\natexlab{a}}){Molteni}, {Ryu}, \&
  {Chakrabarti}}]{mrc1996}
{Molteni}, D., {Ryu}, D., \& {Chakrabarti}, S.~K. 1996{\natexlab{a}}, \apj,
  470, 460, \dodoi{10.1086/177877}

\bibitem[{{Molteni} {et~al.}(1996{\natexlab{b}}){Molteni}, {Sponholz}, \&
  {Chakrabarti}}]{msc1996}
{Molteni}, D., {Sponholz}, H., \& {Chakrabarti}, S.~K. 1996{\natexlab{b}},
  \apj, 457, 805, \dodoi{10.1086/176775}

\bibitem[{{Nishikawa} {et~al.}(2005){Nishikawa}, {Richardson}, {Koide},
  {Shibata}, {Kudoh}, {Hardee}, \& {Fishman}}]{nishi2005}
{Nishikawa}, K.~I., {Richardson}, G., {Koide}, S., {et~al.} 2005, \apj, 625,
  60, \dodoi{10.1086/429360}

\bibitem[{{Paczy{\'n}sky} \& {Wiita}(1980)}]{pw1980}
{Paczy{\'n}sky}, B., \& {Wiita}, P.~J. 1980, \aap, 500, 203

\bibitem[{{Rozyczka} {et~al.}(1996){Rozyczka}, {Bodenheimer}, \&
  {Lin}}]{rbl1996}
{Rozyczka}, M., {Bodenheimer}, P., \& {Lin}, D.~N.~C. 1996, \apj, 459, 371,
  \dodoi{10.1086/176900}

\bibitem[{{Ryu} {et~al.}(1995{\natexlab{a}}){Ryu}, {Brown}, {Ostriker}, \&
  {Loeb}}]{ryu1995apj}
{Ryu}, D., {Brown}, G.~L., {Ostriker}, J.~P., \& {Loeb}, A. 1995{\natexlab{a}},
  \apj, 452, 364, \dodoi{10.1086/176308}

\bibitem[{{Ryu} {et~al.}(1997){Ryu}, {Chakrabarti}, \& {Molteni}}]{rcm1997}
{Ryu}, D., {Chakrabarti}, S.~K., \& {Molteni}, D. 1997, \apj, 474, 378,
  \dodoi{10.1086/303461}

\bibitem[{{Ryu} {et~al.}(1995{\natexlab{b}}){Ryu}, {Yun}, \&
  {Cheo}}]{ryu1995jkas}
{Ryu}, D., {Yun}, H.~S., \& {Cheo}, S.-U. 1995{\natexlab{b}}, Journal of Korean
  Astronomical Society, 28, 223

\bibitem[{{Shafee} {et~al.}(2008){Shafee}, {McKinney}, {Narayan},
  {Tchekhovskoy}, {Gammie}, \& {McClintock}}]{shafee2008}
{Shafee}, R., {McKinney}, J.~C., {Narayan}, R., {et~al.} 2008, \apjl, 687, L25,
  \dodoi{10.1086/593148}

\bibitem[{{Shibata} {et~al.}(1990){Shibata}, {Tajima}, \&
  {Matsumoto}}]{shibata1990}
{Shibata}, K., {Tajima}, T., \& {Matsumoto}, R. 1990, \apj, 350, 295,
  \dodoi{10.1086/168382}

\bibitem[{{Tchekhovskoy} {et~al.}(2011){Tchekhovskoy}, {Narayan}, \&
  {McKinney}}]{tchekhov2011}
{Tchekhovskoy}, A., {Narayan}, R., \& {McKinney}, J.~C. 2011, \mnras, 418, L79,
  \dodoi{10.1111/j.1745-3933.2011.01147.x}

\end{thebibliography}



\end{document}